\documentclass[namedreferences]{SolarPhysics_arxiv}
\usepackage[optionalrh]{spr-sola-addons_arxiv} % For Solar Physics
\usepackage{graphicx}        % For eps figures, newer & more powerfull
\usepackage{color}           % For color text: \color command
\usepackage{url}             % For breaking URLs easily trough lines
\usepackage{txfonts}
\newcommand{\etal}{{\it et al.}}
\newcommand{\eg}{{\it e.g.}}

\newcommand{\Sunrise}{\mbox{\textit{Sunrise} }}

\newcommand{\adv}{    {\it Adv. Space Res.}}

\newcommand{\aap}{    {\it Astron. Astrophys.}}

\newcommand{\apj}{    {\it Astrophys. J.}}

\newcommand{\pasj}{   {\it Pub. Astron. Soc. Japan}}

\newcommand{\solphys}{{\it Solar Phys.}}

\begin{document}

\begin{article}

\begin{opening}

\title{The \Sunrise Mission}

\author{P.~\surname{Barthol}$^{1}$\sep
        A.~\surname{Gandorfer}$^{1}$\sep
        S.~K.~\surname{Solanki}$^{1}$\sep
        M.~\surname{Sch\"{u}ssler}$^{1}$\sep
        B.~\surname{Chares}$^{1}$\sep
        W.~\surname{Curdt}$^{1}$\sep
        W.~\surname{Deutsch}$^{1}$\sep
        A.~\surname{Feller}$^{1}$\sep
        D.~\surname{Germerott}$^{1}$\sep
        B.~\surname{Grauf}$^{1}$\sep
        K.~\surname{Heerlein}$^{1}$\sep
        J.~\surname{Hirzberger}$^{1}$\sep
        M.~\surname{Kolleck}$^{1}$\sep
        R.~\surname{Meller}$^{1}$\sep
        R.~\surname{M\"{u}ller}$^{1}$\sep
        T.~L.~\surname{Riethm\"{u}ller}$^{1}$\sep
        G.~\surname{Tomasch}$^{1}$\sep
        M.~\surname{Kn\"{o}lker}$^{2}$\sep
        B.~W.~\surname{Lites}$^{2}$\sep
        G.~\surname{Card}$^{2}$\sep
        D.~\surname{Elmore}$^{2}$\thanks{Now at National Solar Observatory/Sacramento Peak, P.O. Box 62, Sunspot, NM 88349, USA}\sep
        J.~\surname{Fox}$^{2}$\sep
        A.~\surname{Lecinski}$^{2}$\sep
        P.~\surname{Nelson}$^{2}$\sep
        R.~\surname{Summers}$^{2}$\sep
        A.~\surname{Watt}$^{2}$\sep
        V.~\surname{Mart\'{i}nez~Pillet}$^{3}$\sep
        J.~A.~\surname{Bonet}$^{3}$\sep
        W.~\surname{Schmidt}$^{4}$\sep
        T.~\surname{Berkefeld}$^{4}$\sep
        A.~M.~\surname{Title}$^{5}$\sep
        V.~\surname{Domingo}$^{6}$\sep
        J.~L.~\surname{Gasent~Blesa}$^{6}$\sep
        J.~C.~\surname{del~Toro~Iniesta}$^{7}$\sep
        A.~\surname{L\'{o}pez~Jim\'{e}nez}$^{7}$\sep
        A.~\surname{\'{A}lvarez-Herrero}$^{8}$\sep
        L.~\surname{Sabau-Graziati}$^{8}$\sep
        C.~\surname{Widani}$^{9}$\sep
        P.~\surname{Haberler}$^{9}$\sep
        K.~\surname{H\"{a}rtel}$^{9}$\sep
        D.~\surname{Kampf}$^{9}$\sep
        T.~\surname{Levin}$^{9}$\sep
        I.~\surname{P\'{e}rez Grande}$^{10}$\sep
        A.~\surname{Sanz-Andr\'{e}s}$^{10}$\sep
        E.~\surname{Schmidt}$^{11}$
       }
\runningauthor{Barthol et al.}
\runningtitle{The \Sunrise Mission}

   \institute{$^{1}$ Max-Planck-Institut f\"{u}r Sonnensystemforschung, Max-Planck-Stra\ss e~2, D-37191~Katlenburg-Lindau, Germany \\
                     email: \url{barthol@mps.mpg.de} \\
              $^{2}$ High Altitude Observatory\thanks{HAO/NCAR is sponsored by the National Science Foundation}, P.O. Box 3000, Boulder, CO~80301, USA \\
%                     email: \url{e.mail-c} \\
              $^{3}$ Instituto de Astrof\'{i}sica de Canarias, C/V\'{i}a L\'{a}ctea, s/n, E-38205, La~Laguna (Tenerife), Spain \\
%                     email: \url{e.mail-c} \\
              $^{4}$ Kiepenheuer-Institut f\"{u}r Sonnenphysik, Sch\"{o}neckstra\ss e~6, D-79104~Freiburg, Germany \\
%                     email: \url{e.mail-c} \\
              $^{5}$ Lockheed Martin Solar and Astrophysics Laboratory, Bldg.~252, 3251~Hanover~Street, Palo~Alto, CA~94304, USA \\
%                     email: \url{e.mail-c} \\
              $^{6}$ Grupo de Astronom\'{i}a y Ciencias del Espacio (Univ.~de~Valencia), E-46980, Paterna, Valencia, Spain \\
%                     email: \url{e.mail-c} \\
              $^{7}$ Instituto de Astrof\'{i}sica de Andaluc\'{i}a (CSIC), Apdo.~de~Correos 3004, E-18080, Granada, Spain \\
%                     email: \url{e.mail-c} \\
              $^{8}$ Instituto Nacional de T\'{e}cnica Aerospacial, E-28850, Torrej\'{o}n~de~Ardoz, Spain \\
%                     email: \url{e.mail-c} \\
              $^{9}$ Kayser-Threde GmbH, Wolfratshauser Stra\ss e~48, D-81379~M\"{u}nchen, Germany \\
%                     email: \url{e.mail-c} \\
              $^{10}$ Universidad Polit\'{e}cnica de Madrid, IDR/UPM, Plaza Cardenal Cisneros~3, E-28040, Madrid, Spain \\
%                     email: \url{e.mail-c} \\
              $^{11}$ Ingenieurb\"{u}ro f\"{u}r Optikentwicklung, Amalienstra\ss e~12, D-85737~Ismaning, Germany \\
%                     email: \url{e.mail-c} \\
             }

\begin{abstract}

The first science flight of the balloon-borne \Sunrise telescope
took place in June 2009 from ESRANGE (near Kiruna/Sweden) to Somerset
Island in northern Canada. We describe the scientific aims and mission
concept of the project and give an overview and a description of the
various hardware components: the 1-m main telescope with
its postfocus science instruments (the UV filter imager SuFI and the
imaging vector magnetograph IMaX) and support instruments (image
stabilizing and light distribution system ISLiD and correlating wavefront
sensor CWS), the optomechanical support structure and the
instrument mounting concept, the gondola structure and the power,
pointing, and telemetry systems, and the general electronics architecture.
We also explain the optimization of the structural and thermal design
of the complete payload. The preparations for
the science flight are described, including AIV and ground calibration
of the instruments. The course of events during the science
flight is outlined, up to the recovery activities. Finally, the in-flight performance
of the instrumentation is briefly summarized.

\end{abstract}
\end{opening}

%-------------------------------------------------

\section{Introduction}
     \label{S-Introduction}
\subsection{Solar Balloon Missions}

A number of projects has been devoted in the past to observing the Sun from
balloon-borne platforms in the stratosphere. In 1957 and 1959, project {\em
Stratoscope} (Schwarz\-schild, 1959) obtained white-light images of granulation
and sunspots with a 30-cm telescope (Danielson, 1961). A different instrument
called {\em Spektro-Stratoskop} took several long sequences of high quality
granulation pictures with an evacuated 30-cm telescope (Mehl\-tretter, 1978).
From 1966 to 1973, there were several flights of the {\em Soviet
Stratospheric Solar Station}, which operated alternatively with main mirrors
of 50 cm and 1 m diameter and took filtergrams in the visible range (\eg
Krat \etal, 1972). Smaller telescopes (5--10~cm) with broad-band filters in
the visible were flown by a Japanese group in the 1970s (Hirayama, 1978).
Groups in France attempted UV imaging in the wavelength range 200--300~nm
with a 20-cm telescope (Herse, 1979) and took spectra in the same wavelength
range with telescopes of up to 30~cm aperture from the mid 1960s on
(Samain and Lemaire, 1967; Lemaire and Blamont, 1985). More recently, flights
of the {\em Flare Genesis} project in 1996 and 2000 provided magnetograms and
Dopplergrams in the CaI 6122.2 line using a 80-cm telescope (Rust \etal,
1996; Bernasconi \etal, 2000). The 30-cm telescope of the {\em Solar
Bolometric Imager} (Bernasconi \etal, 2004) provided maps of the solar disk
in integrated light between 0.28 and 2.6~$\mu$m during three flights between
2003 and 2007.

These projects demonstrated the advantages and the rich potential
of balloon-borne solar observations at stratospheric heights: (i)
negligible image degradation by atmospheric seeing, (ii) access to the UV
range down to about 200~nm wavelength, (iii) retrievability of the
instruments, and (iv) much reduced cost in comparison to space projects.

The \Sunrise mission comprises the biggest and most complex
payload flown so far in a solar balloon mission: a 1~m telescope
equipped with a multi-wavelength UV filter imager, a Fabry-P\'erot-based
vector magnetograph, and a correlating wavefront sensor for active
alignment control and image stabilization.

\subsection{Science Case}

The solar photosphere is crucial for the investigation of the solar magnetic
field. This thin layer, where the plasma becomes optically thin and almost
all of the radiative energy flux is emitted, represents the key interaction
region: thermal, kinetic and magnetic energy all are of the same order of
magnitude and transform most easily from one form into another. The
interaction between convection, radiation, and magnetic field in the
electrically conducting solar plasma leads to the creation of a rich variety
of magnetic structure with intense (kilogauss) magnetic field concentrations
on size scales reaching well below a pressure scale height. At the same time,
the photosphere appears to harbor a stunning amount of mixed-polarity
`turbulent' magnetic flux, which possibly results from small-scale local
dynamo action driven by the granulation flows. The photospheric magnetic
field is in a state of constant change: ceaseless transport, stretching,
emergence and submergence, cancellation, intensification and dissipation of
magnetic flux take place down to the smallest spatial scales that can be
observed so far. These processes control the structure, dynamics, and
energetics of the solar atmosphere at larger scales; they are the source of
solar variability and, ultimately, of solar influences on the Earth. To
understand these fundamental processes, we must learn how the magnetic field
interacts with the solar plasma and have to uncover the conversion of energy
between its mechanical, magnetic, radiative and thermal forms. Consequently,
the central questions of \Sunrise science are %
\begin{itemize}
\item What are the origin and the properties of the intermittent
      magnetic structure, including the kilogauss concentrations?
\item How is the magnetic flux brought to and removed from the solar
      surface? What is the role played by local dynamo action and
      reconnection processes?
\item How does the magnetic field assimilate and provide energy to heat
      the upper solar atmosphere?
\item How does the variable magnetic field modify the solar brightness?
\end{itemize}

In order to answer these questions, the key scientific objective of the
\Sunrise project is to study the structure and dynamics of the solar magnetic
field at the spatial resolution afforded by a 1-m telescope, over extended
stretches of time with constant observational conditions free of seeing
effects, with a time resolution sufficient to track rapid changes of the
magnetic field, and over a field of view large enough to provide good
statistics of relevant events and to follow the evolution of magnetic
structure during all phases of its life cycle.

The required observations include continuous, quantitative measurements of
the \linebreak [4] magnetic field vector, the plasma velocity and related
atmospheric structure, together with brightness maps in several wavelength
bands. For the latter, the UV wavelengths between 200~nm and 300~nm are of
particular interest since a) no high-resolution imaging has been done so far
in this range and b) cycle-related irradiance variations in this spectral
band affect the temperature in the terrestrial stratosphere through Ozone
photochemistry. Numerical simulations predict intensity contrasts of magnetic
flux concentrations at these wavelengths that significantly exceed those in
the visible.

\subsection{Mission Concept}

\Sunrise has been designed as a balloon-borne stratospheric solar observatory
in the framework of NASA's LDB (Long Duration Balloon) program. Zero pressure
helium balloons with volumes up to 1~million cubic meters lift science
payloads such as \linebreak[4] \Sunrise with a mass of several tons to float altitudes of
35--37~km. This 'low cost access to near space' has several advantages
compared to ground-based or space-borne solar instrumentation. As with
space-borne instruments, stratospheric solar observations do not suffer from
two major drawbacks encountered with ground based telescopes. Being above
99~\% of the Earth's atmosphere, wavefront distortions due to atmospheric
turbulence are virtually not existent. This provides long-term and
seeing-free high resolution imaging conditions. In addition, the interesting
solar UV radiation between 220~nm and 370~nm is accessible, which is absorbed
in the lower stratosphere by the Hartley and Huggins bands of Ozone. At
mission termination balloon-borne instruments usually can be recovered after
landing with only moderate damage. Modifications for a future flight and
refurbishment can be accomplished very cost effectively, at far lower
investment levels compared to space instruments.

NASA LDB missions are managed and operated by the Columbia Scientific
Ballooning Facility (CSBF), Palestine, Texas, USA and in the past have been
launched mainly from Williams Field near McMurdo, Antarctica
(77.86$^\circ$~S, 167.13$^\circ$~E), with a launch window from December to
January. NASA recently expanded their launch capabilities by cooperating with
ESRANGE (67.89$^\circ$~N, 21.10$^\circ$~E) near Kiruna, Sweden, allowing
missions to be launched during the northern summer period as well. The
European sounding rocket and balloon base provides excellent infrastructure
and is accessible with much lower logistical effort compared to McMurdo.
Launching balloons during solstice conditions close to the polar circle
offers uninterrupted solar observations without day/night cycles. Permanent
sunlight and only small elevation changes of the Sun form ideal conditions,
so that undisturbed observation and power generation for the instruments are
guaranteed. Furthermore, thermal conditions do not vary significantly and the
balloon floats at nearly constant altitude. Relatively stable wind systems at
float altitude take the balloon and instrument on circumpolar trajectories
with flight durations of 9 to 12 days per revolution. ESRANGE was chosen for
the first science flight of \Sunrise, although the flight duration currently
is limited to 5--6~days due to the lack of Russian overflight permissions for
NASA balloons. The processes to be studied by \Sunrise generate highly
variable structures like emerging and decaying active regions or sunspots.
The timescales involved range from seconds on small spatial scales over
minutes and hours to several days in case of sunspots. Due to their intrinsic
nature these phenomena cannot be predicted in time and location. Being at the
onset of the new solar activity cycle, the shorter flight duration was judged
to be acceptable in comparison to the higher logistical effort and risk
related with an Antarctic flight.

Coordinated measurements were done with HINODE and SUMER onboard SOHO.
Parallel observations with ground-based solar telescopes such as the Swedish
Solar Telescope (SST) on La Palma, the Vacuum Tower Telescope (VTT) on
Tenerife, the IRSOL facility in Locarno (Switzerland), and the Dunn Solar
Telescope on Sacramento Peak (New Mexico) have been performed to support the
target selection for \Sunrise.

\section{Instrument Description}

\begin{figure*}
 \centerline{\includegraphics[width=1.\textwidth,clip=]{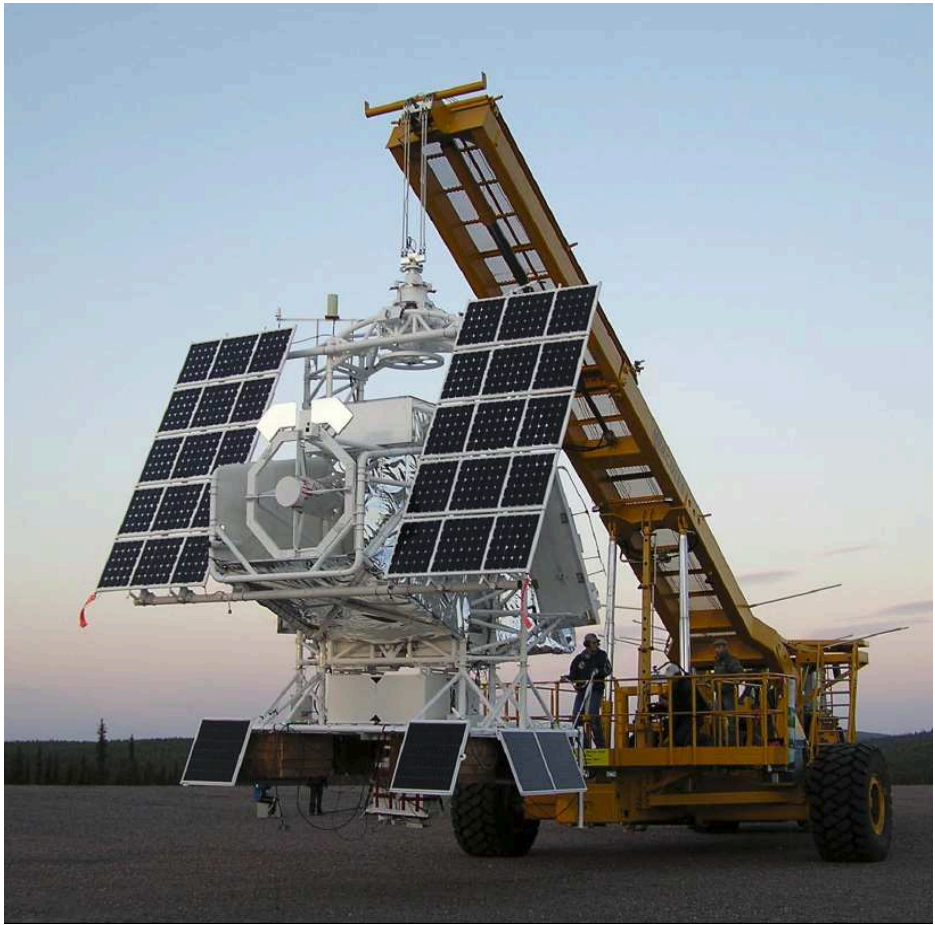}}
\caption{Overview of the \Sunrise observatory in full flight configuration,
immediately before launch. The dimensions of the observatory can be judged by
comparing with the person on the launch vehicle. In the center of the gondola
the telescope front ring can be seen, with the radiators for the heat
rejecting prime focus field stop. The box on top of the telescope is the
postfocus instrumentation (PFI) platform housing the science instruments. A
flywheel below the upper bridge of the gondola structure is used for azimuth
pointing. Instrument electronics is mounted on two inclined racks on the rear
side of the gondola. The white box in the lower part of the gondola is the
CSBF-provided System Instrumentation Package (SIP), needed for commanding of
balloon and payload. It uses a dedicated power system, while the main power
for the observatory is generated by the large solar arrays to both sides of
the telescope. }
\label{overview}
\end{figure*}

\Sunrise in full flight configuration is shown in Figure~\ref{overview}. The
instrument can be broken down into the following major components, which will
be described in more detail below:

\begin{itemize}
\item Telescope
\item Postfocus instrumentation platform (PFI) with science instruments and image
stabilization system
\item Gondola with photovoltaic arrays and pointing system
\item  Instrument electronics
\item  CSBF-provided ballooning equipment
\item Ground-support equipment
\end{itemize}

\subsection{Telescope}

\begin{figure*}[t]
\centerline{\includegraphics[width=1.\textwidth,clip=]{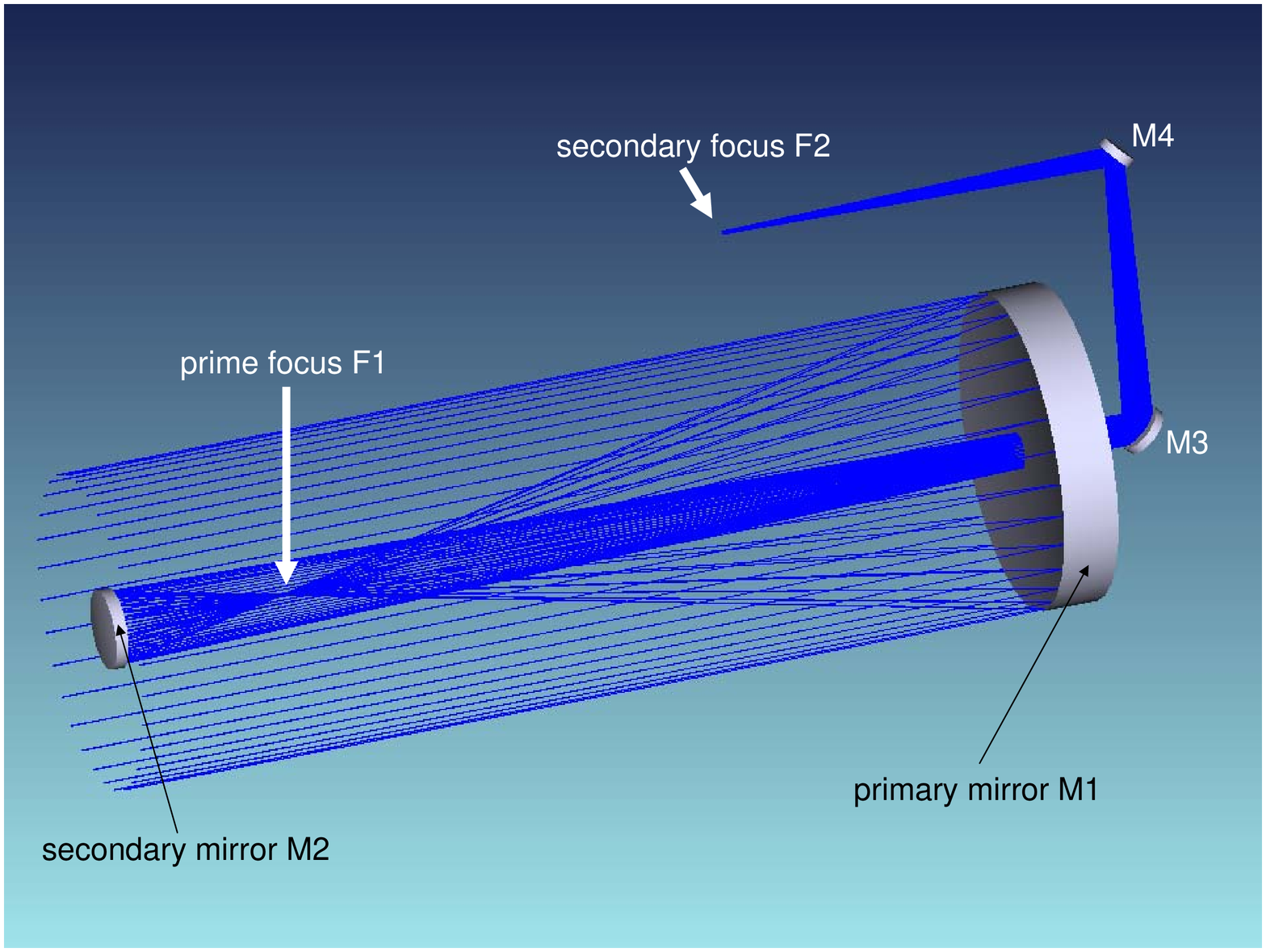}}
\caption{Optical configuration of the 1~m Gregory telescope of \Sunrise. }
\label{telescope_path}
 \end{figure*}

The \Sunrise telescope is a light-weight Gregory-type reflector with nearly
25~m effective focal length. The optical path is depicted in
Figure~\ref{telescope_path}. A parabolic primary mirror with 1010~mm outer
diameter and 1000~mm clear aperture creates a real image of the sun in the
first focus F1 of the telescope, 2422.9~mm in front of the M1 vertex. A field
stop in F1 limits the telescope field-of-view (FoV) so that only a small
fraction reaches the elliptical secondary mirror M2. M2 has a diameter of
245~mm and a focal length of 505~mm. M2 magnifies the image in F1 by a factor
of ten and refocuses the radiation to a secondary telescope focus, located in
the postfocus instrumentation package ontop the central telescope frame. Two
plane fold mirrors M3 and M4 redirect the light from the telescope to this
upper level. A further physical stop in the secondary focus limits the
telescope FoV to 180~arcsec.

\begin{figure*}[t]
 \centerline{\includegraphics[width=1.\textwidth,clip=]{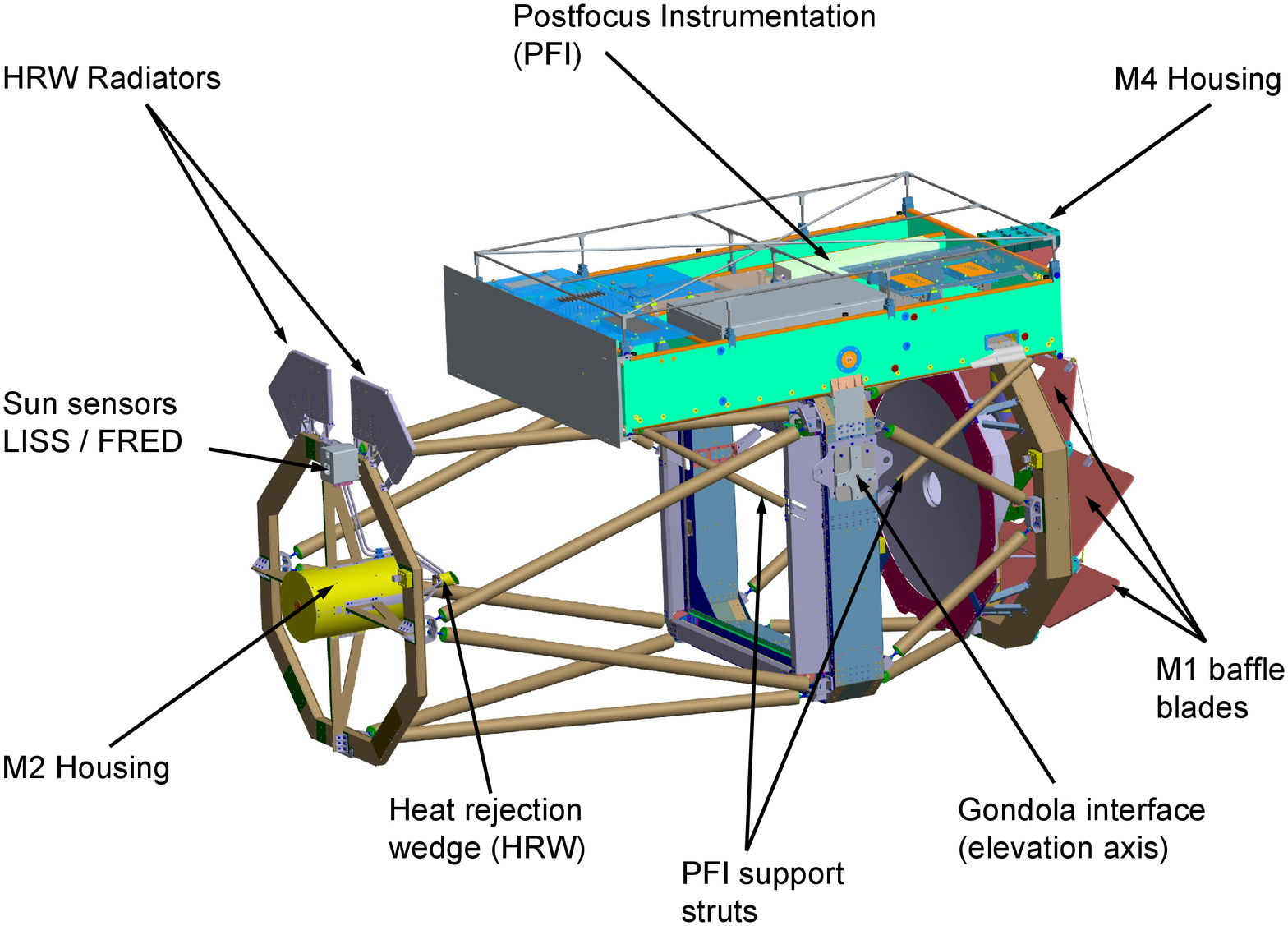}}
\caption{Overall sketch of the telescope Serrurier structure and
subassemblies} \label{telescope_CAD}
 \end{figure*}

\begin{figure*}[t]
 \centerline{\includegraphics[width=1.\textwidth,clip=]{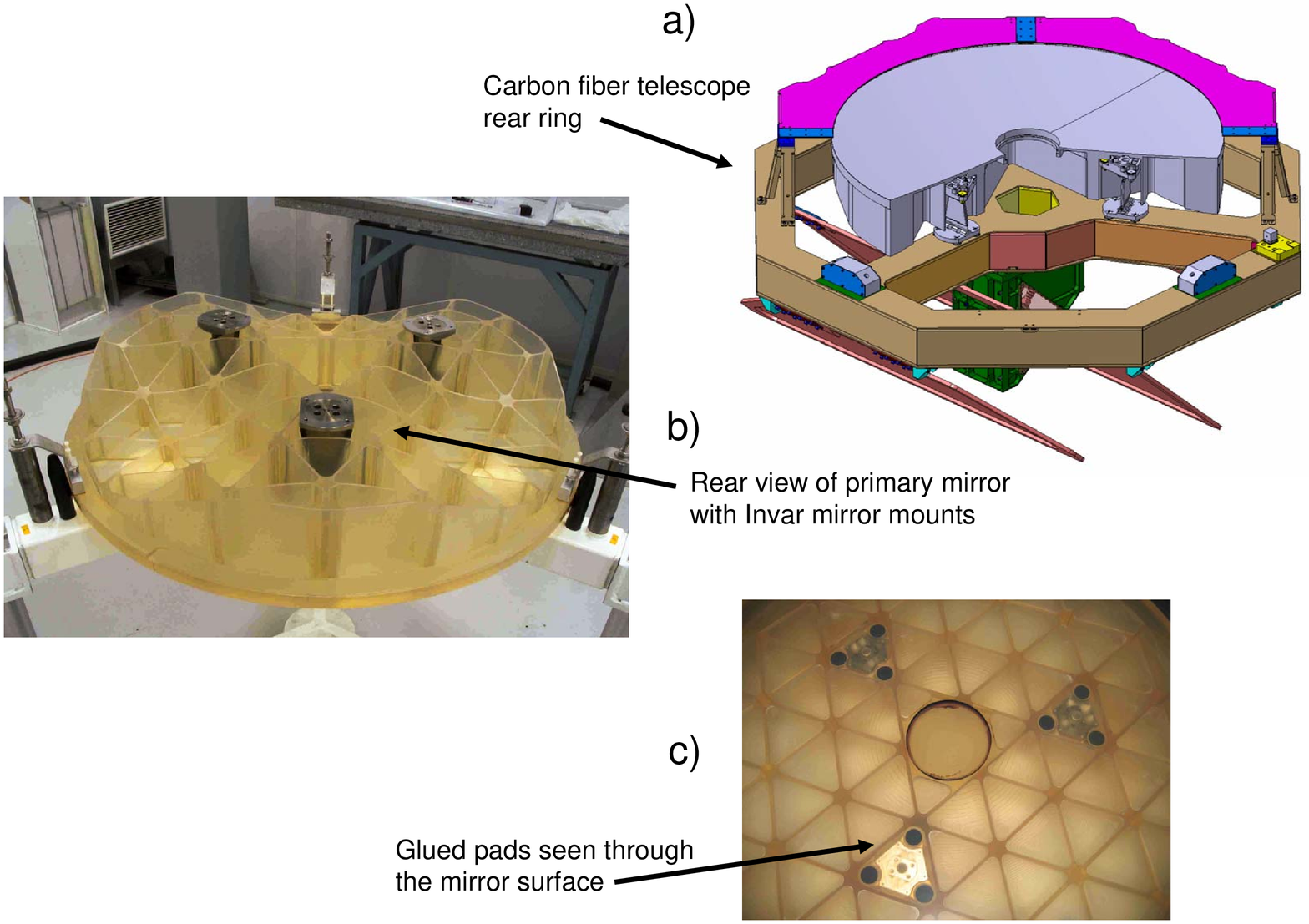}}
\caption{a) CAD model of the main mirror cell with the lightweighted ZERODUR
mirror. The mirror is mounted with three flex blade fixation points to the
carbon fiber structure, which forms the rear ring of the telescope. b)
Backside view of main mirror after integration of the three flex blade
fixation points. c) Glass/metal junction formed by glued Invar pads }
\label{mirror_CAD}
 \end{figure*}

The Serrurier structure (Figure~\ref{telescope_CAD}) connects the steel central
frame, the front ring with the secondary mirror assembly and heat rejection
wedge, and the rear ring with the primary and tertiary mirrors. The central
frame serves as the main structural component. It is a riveted sheet metal
construction with metallic inserts at locations with high mechanical load.
The frame carries the postfocus instrumentation (PFI) piggy-back on top.
Axles mounted left and right to interface plates at the height of the
combined telescope and PFI center-of-gravity form the elevation axis that
connects to the gondola structure. The front ring carries the housing of the
secondary mirror assembly and the heat rejection wedge including radiators.
On its front side facing towards the Sun, two Sun sensors are mounted. The
LISS (Lockheed Intermediate Sun Sensor) and Full Range Elevation Device
(FRED) are part of the gondola pointing and tracking control system (Section
2.3.3), with the LISS serving as pointing reference in flight. The back ring
(see also Figure~\ref{mirror_CAD}a) provides a stiff platform for the primary
mirror and aperture stop. On its rear side, the tertiary plane fold mirror is
mounted, as well as three blades providing thermal control of the primary
mirror. Front and rear ring are connected to the central frame with eight
struts each, forming the characteristic Serrurier mounting. This mounting
maintains parallelism of the main elements even in case of relative lateral
displacements due to changing gravity loads. The rings as well as the
connecting struts are made of carbon fiber reinforced plastic composite
materials for high stiffness, low weight and low thermal expansion. The
stiffness of the struts is designed such that front and rear ring show
approximately the same lateral displacement in the presence of gravitational
loads, keeping the relative positioning of M1 and M2 within fractions of a
millimeter. The connection of the struts to the central frame was made by
steel shafts, which allowed easy and quick dismantling of the main telescope
parts in the field after recovery.

The main mirror has been manufactured by SAGEM in France. Schott grade 0
Zerodur is used as substrate material. The rear side of the mirror has a
three-fold symmetry with triangular honeycomb structuring, optimizing
stiffness under varying gravitational loads (Figure~\ref{mirror_CAD}b). Its
thickness is 178~mm with a weight of less than 47~kg. This extreme
lightweighting is achieved by mechanical milling and acid etching. Three
Invar mounts connect the mirror with the rear ring of the telescope structure
(Figure~\ref{mirror_CAD}a). On each mount, the glass-metal junction is formed
by 3~Invar pads of 25~mm diameter, which are glued to the structure behind
the face sheet (Figure~\ref{mirror_CAD}c). The facesheet of the mirror was
ground and polished in a conventional way. As final surface treatment step,
ion-beam figuring was applied. This technique removes residual quilting,
which is unavoidable when manufacturing lightweighted mirrors with facesheet
thicknesses in the order of 7~mm. To further improve the overall wavefront
quality, a dedicated map-imprint was applied during the ion-beam figuring.
Based on finite-element mechanical deformation calculations and analysis of
the various thermal load cases, a compromise was calculated providing best
wavefront error (WFE) performance at 22.5$^\circ$ elevation, with only
minimal degradation in the range from 0$^\circ$ to 45$^\circ$ elevation
angle, which is the nominal elevation range of \Sunrise. The measured WFE
(@633~nm) at 0$^\circ$ elevation is only 30.2~nm rms, reducing to 19.2~nm rms
at 22.5$^\circ$ and 24.1~nm rms at 45$^\circ$ elevation. The reflective
coating on the front face is bare aluminum with a thickness of 100~nm. An
aperture stop with 1000 mm~inner diameter directly in front of M1 defines the
entrance pupil of the system.

\begin{figure*}[t]
 \centerline{\includegraphics[width=1.\textwidth,clip=]{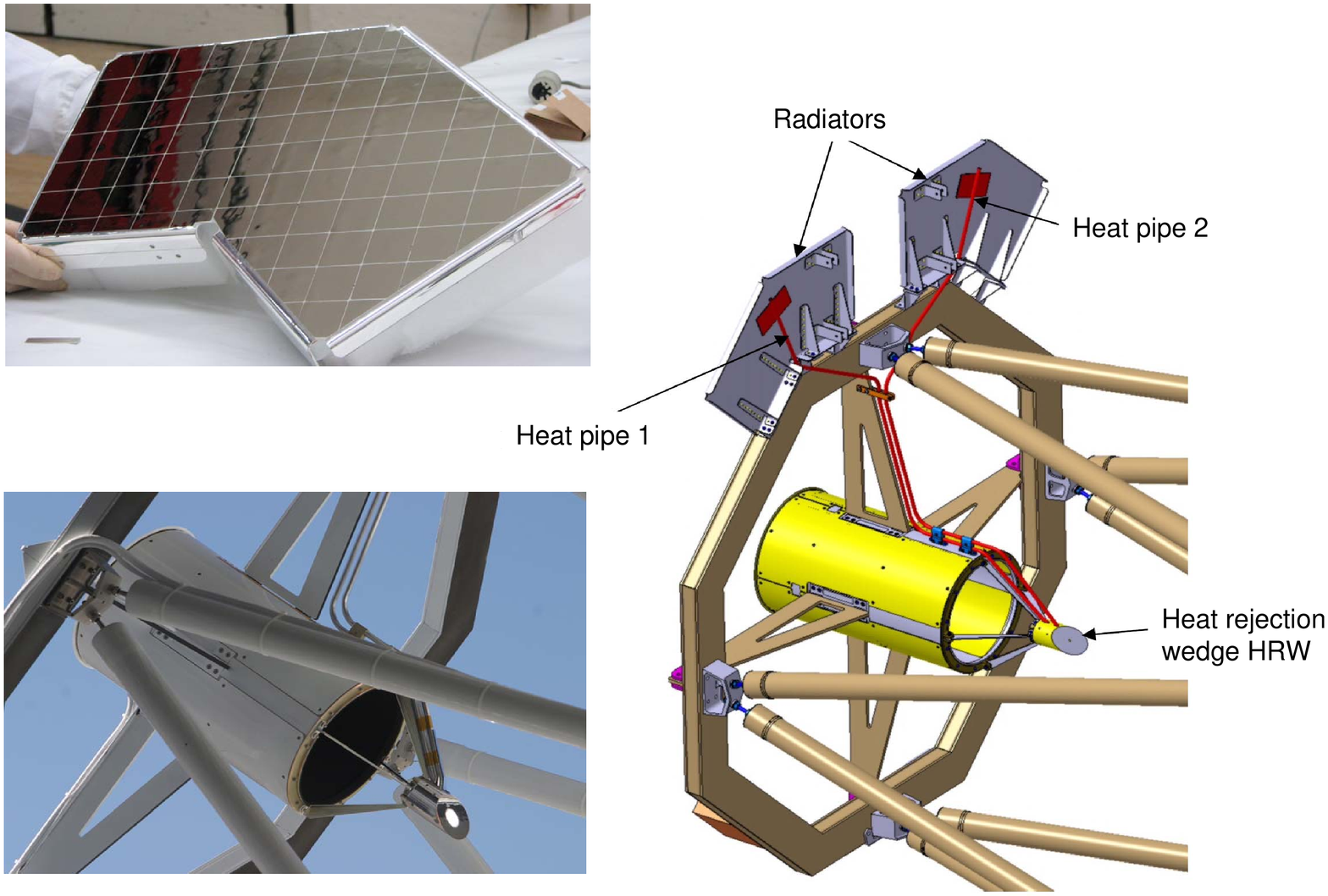}}
\caption{CAD model of the primary field stop (heat rejection wedge, HRW) with
the two heat pipes, which drain the absorbed energy to the two radiators on
top of the front ring. The radiators are facing direct sunlight and are
therefore covered with Optical Solar Reflectors (OSRs), thin second surface
mirrors with high thermal emissivity, as shown in the insert on the top left.
The bottom left picture shows the image of the solar disk on the HRW during
ground testing. } \label{HRW}
 \end{figure*}

The primary mirror creates a real image of the sun in the primary focus F1 of
the telescope. Nearly 1~kW solar radiation is concentrated on a disk of about
22~mm diameter. At this position a heat rejection wedge (HRW) with a central
hole acts as field stop (Figure~\ref{HRW}).

The 2.8~mm hole -- and a slightly undersized stop in the conjugate image
plane in F2 -- define the useable telescope field-of-view (FoV),
corresponding to 130~Mm on the solar surface. The HRW is a highly conductive
cylindric aluminum block with a wedged front face reflecting 99 \% of the
incoming light out of the telescope. The small fraction of solar radiation
passing through the field stop reduces the heat load on the science
instrumentation to about 10~W. Absorption of the focused energy at the HRW is
minimized by a second surface mirror, glued to the aluminum body. A thin
glass plate with vapour deposited aluminum on the rear side acts as optical
solar reflector; UV reflectivity is enhanced by a dedicated coating on its
front surface. The size of the HRW is about twice the diameter of the solar
disk image, thus allowing solar limb observations, while keeping the solar
image on the HRW. Two ammonia heat pipes connect the HRW to dedicated
radiators. Equipped with optical solar reflectors -- same second surface
mirrors, but without UV coating -- and pointing towards the Sun, the
radiators efficiently cool the HRW to temperatures less than 25$^\circ$C,
thus avoiding any Schlieren build-up, which could cause wavefront
distortions.

The optical system of the \Sunrise telescope is semi-active in order to
maintain the highest performance throughout the flight. The secondary mirror
M2, polished and ion-beam figured to a residual WFE below 6~nm rms, is
isostatically mounted on bipods and connected to a three-axis translation
stage. The M2 position is fine adjustable in a range of 1.2~mm to an accuracy
of 5~$\mu$m laterally and 1~$\mu$m axially, so that the relative M1/M2
alignment can be kept constant even under varying telescope elevation and
thermal loads. A wavefront sensor located in the postfocus instrumentation
(see below) monitors the alignment status and generates control signals for
M2 mirror re-positioning. The secondary mirror assembly with its housing
creates a central obscuration of 324~mm diameter for the telescope.

The folding mirrors M3 and M4 are equipped with translation stages. It was
originally foreseen to use them for fine focussing. However, the adjustment
accuracy of the M2 axial translation stage proved to be sufficiently high, so
that focussing could be performed with M2. The WFE increase due to spherical
aberration when axially moving M2 is negligible. M3 and M4 are only adjusted
during static alignment on ground, determining beam height and secondary
focus position within the PFI as well as the lateral position of the pupil
image at the image stabilization tip-/tilt mirror.

Thermal control of the primary mirror is essential for the performance of the
telescope. About 80~W solar radiation are absorbed in the coating. Dedicated
baffle blades behind the mirror with reflective front sides increase the view
factor to the cold sky and shade the mirror against Earths IR radiation and
reflected sunlight from the ground (ice) or sea below the balloon.

The energy density in the primary focus is high enough to damage structural
parts in case of uncontrolled beam wandering. A retractable curtain in the
plane of the central frame can close the rear compartment of the telescope in
case of pointing loss, acting as an aperture door. The curtain needs about 20
seconds to securely block the telescope aperture. This is well below critical
exposure times of parts in the vicinity of the first field stop. In-flight
operation of the curtain is controlled by two software flags set by the
pointing system computer. Automatic closure is initiated when the pointing
error exceeds $\pm$15~arcmin (coarse pointing flag 'off', solar disk image
could leave heat rejection wedge). Re-opening is initiated if the pointing
remains stable within a cone of 20~arcsec (fine pointing flag 'on', pointing
is within range of image stabilization).

The telescope has been built by Kayser-Threde, Munich, under contract of MPS.

\subsection{Postfocus Instrumentation Platform}

\begin{figure}[t]
\centerline{\includegraphics[width=1.\textwidth,clip=]{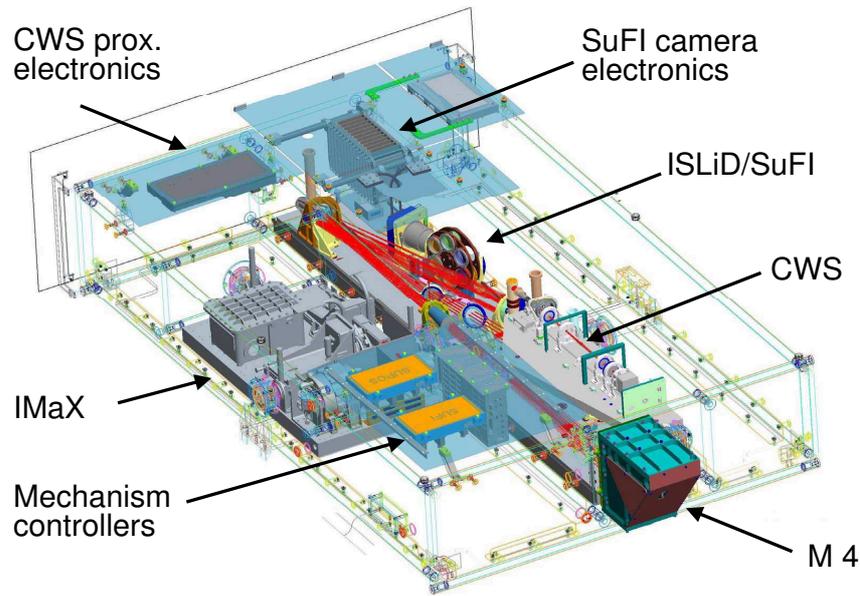}}
\caption{Semi-transparent view of the \Sunrise postfocus instrumentation. Instrument covers are removed.}
\label{PFI}
\end{figure}

The \Sunrise Post-Focus Instrumentation (PFI) rests piggy-back ontop the
telescope. The compact package consists of a rigid support structure and four
instrument modules with their supporting proximity electronics
(Figure~\ref{PFI}). Two of the four instrument modules are ''service'' units:
the Image Stabilization and Light Distribution (ISLiD) and Correlation
tracking and Wavefront Sensing (CWS) units. The science instrumentation
consists of the \Sunrise Filter Imager (SuFI), and the Imaging Magnetograph
Exeriment (IMaX). Mechanism controllers for ISLiD and SuFI as well as
proximity electronics and power supplies for CWS and SuFI are located inside
some of the PFI compartments. The PFI structure provides room for an
additional 3rd science instrument for future flights.

\subsubsection{Image Stabilization and Light Distribution System: ISLiD}

\begin{figure}[t]
\centerline{\includegraphics[width=1.\textwidth,clip=]{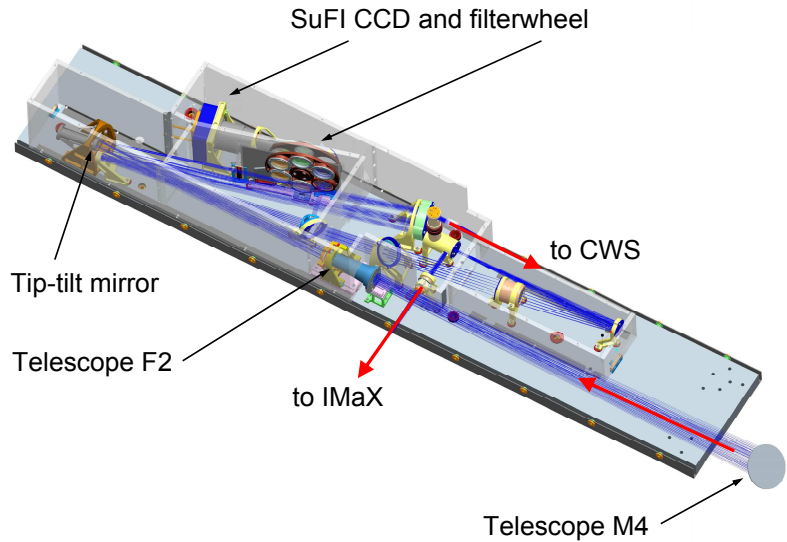}}
\caption{The Image Stabilization and Light Distribution unit (ISLiD) is mounted into the central compartment of the PFI structure.
It distributes the light coming from the telescope (directions shown in red) onto the science instruments and the CWS.}
\label{ISLID_path}
\end{figure}

\begin{figure}[t]
\centerline{\includegraphics[width=1.\textwidth,clip=]{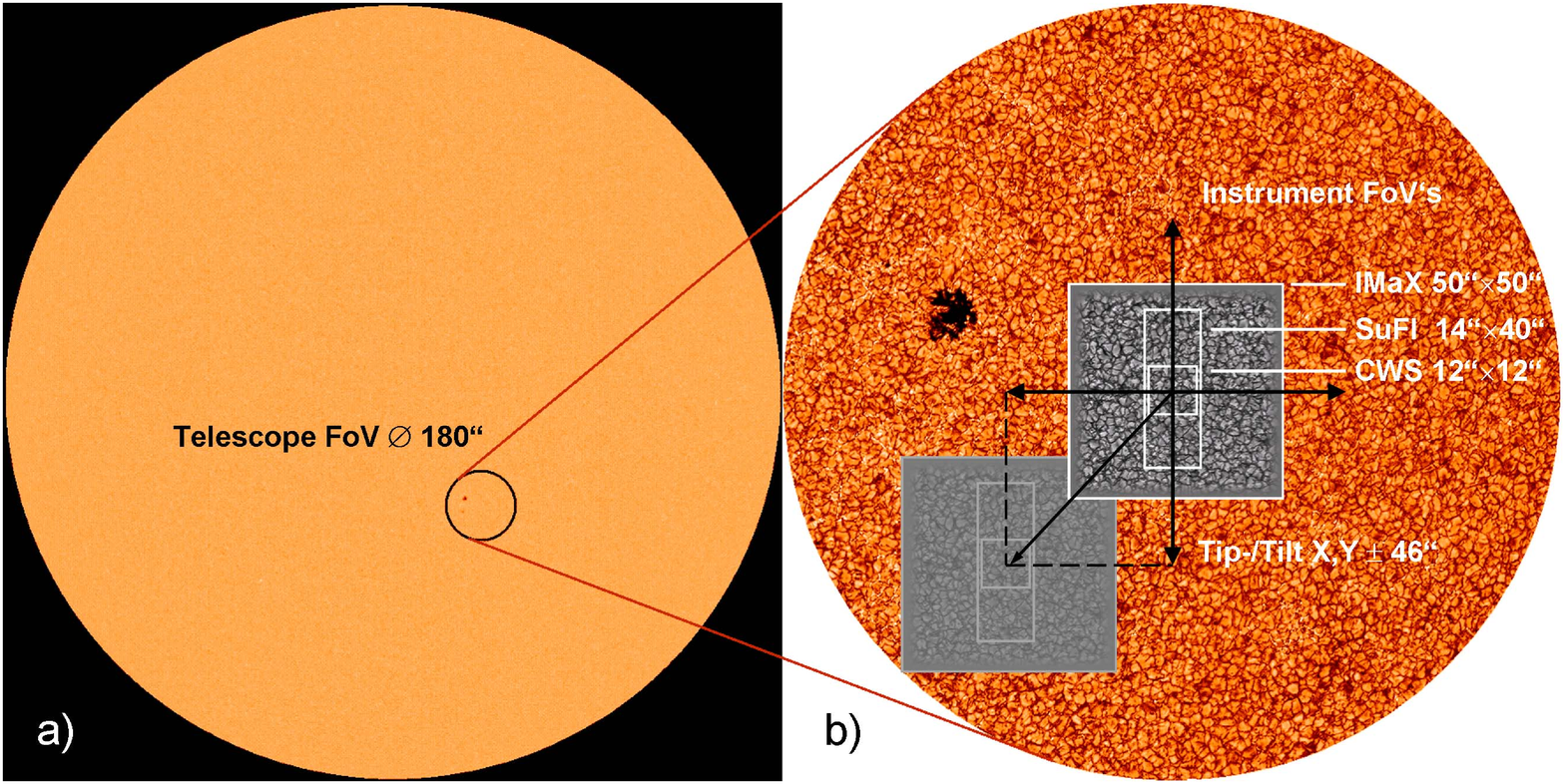}}
\caption{Telescope and instrument fields-of-view (FoV). a) Continuum image of
the solar disk. A circle indicates the \Sunrise telescope FoV of 180~arcsec
diameter, corresponding to roughly 10\% of the solar disk diameter. b)
Co-aligned instrument FoV's of IMaX, SuFI and CWS within the telescope FoV.
The free range of the image stabilization system ($\pm$46~arcsec) is indicated. }
\label{FOV}
\end{figure}

The \Sunrise science requirements demand precision fine pointing and
simultaneous observations by all science instruments. This is ensured by
ISLiD, a panchromatic reimager based on dichroic beamsplitters, which provide
the different wavelength bands to the individual science branches with
maximum photon flux, while preserving the polarization of the incoming light.
ISLiD is located in the center of the PFI structure and takes up its full
length (Figure~\ref{ISLID_path}). Reimaging of the secondary telescope focus
onto the instrument detectors is achieved with a two mirror Schwarzschild
arrangement, before separating the ultraviolet below 400~nm towards SuFI.
Additional refractive optics provide image scaling and telecentricity for
IMaX and CWS, which is fed by light outside the spectral bands used by the
science instruments. ISLiD contains a fast piezo-driven tip-tilt mirror at a
pupil plane of the optical system. A field lens in the telescope secondary
focus projects the aperture stop of the primary mirror onto the tip-tilt
mirror. In order to allow for the UV part of the solar spectrum to be
transmitted, the field lens is made from fused silica and uncoated. The
tip-tilt mirror is part of the CWS system (see below) and used to compensate
for residual image motion due to solar rotation or gondola shake and
vibrations within the instrument. The sizing of telescope and instrument
FoV's allows a capture range for the image stabilization system of
$\pm$46~arcsec (Figure~\ref{FOV}), still avoiding vignetting instrument FoV's
by the secondary field stop in F2. Details of ISLiD (developed by MPS) are
given by Gandorfer \etal, 2010. The image stabilization system and its
damping characteristics are described in Berkefeld \etal, 2010 and in the
following section.

\subsubsection{Correlating Wavefront Sensor: CWS}

The CWS, a Shack-Hartmann type wave front sensor, is located close to ISLiD
in the central compartment of the PFI structure. The high speed camera system
has a FoV of 12$\times$12 arcsec$^2$ on the sky. The CWS is used for two
purposes, for (high frequency) precision image stabilization and guiding and
to control the alignment of the telescope (low frequency). A lenslet array in
a pupil plane has 6 subapertures arranged in a concentric ring, forming 6
independent images on the detector. The information derived from the 6
independently analyzed images of the same solar scene are used to measure the
local wavefront tilt per subaperture, providing the coefficients of a Zernike
polynomial decomposition of the wavefront error up to the third radial
degree. The coefficients for tip and tilt, defocus, and Seidel coma are used
as error signals. A control loop time-integrates these error signals and
converts them into actuation signals to drive the telescope secondary mirror
M2. The fast read-out of the CWS camera ($>$1~kHz) allows detecting
correlated image motion of the 6 separately generated images on the detector
caused by residual uncompensated gondola movements and vibration as well as
by the slow drift of solar features due to solar rotation. Fast software
routines convert the correlation signals to actuator signals for the tip-tilt
mirror, performing the pointing correction, image stabilization and guiding.
The closed-loop control of the image stabilization system has a bandwidth of
about 60 Hz (at the 6dB level) with a sensitivity of better than
0.003~arcsec. It provides very efficient damping of low frequencies
($\sim$600 at 2~Hz, $\sim$70 at 10~Hz), where the pointing control cannot be
provided solely by the gondola. The CWS was activated whenever the gondola
pointing was within its nominal range (see Section 2.3.3.). The CWS unit
including the tip-tilt mirror, electronics and the control software were
developed by the Kiepenheuer-Institut f\"{u}r Sonnenphysik, Freiburg,
Germany. Details can be found in Berkefeld \etal, 2010.

\subsubsection{\Sunrise Filter Imager: SuFI}

SuFI provides high-resolution images of photospheric and chromospheric
features at 5 wavelength bands in the visible and near ultraviolet at high
cadence. The optical arrangement is a Schwarzschild 2-mirror system (part of
ISLiD), magnifying the telescope secondary focus by a factor of 5 onto the
detector, thus resulting in an effective focal length of 121~m. The sensor is
a 2k by 2k UV enhanced fast read-out CCD. The field-of-view of the instrument
is 15$\times$40~arcsec$^2$. Interference filters are employed to isolate the
observed wavelength band. They have peak transmissions at wavelengths of 214
(10)~nm and 300 (5)~nm together with OH and CN molecular absorption bands at
312 (1)~nm and 388 (0.8)~nm and the Ca II H line (singly ionized Calcium) at
397.6 (0.18)~nm. The numbers in brackets are the filter FWHM. In order to
achieve near diffraction-limited imaging, a phase-diversity imaging technique
is used by splitting the image in front of the CCD detector: half of the CCD
area collects the focused image, while an optical delay line forms a second
image with a defocus of approximately one wave on the other half of the CCD.
Post-flight restoration of the images free from static aberrations of the
optical path can thus be achieved. The SuFI optical elements, mechanisms and
CCD are mounted side-by-side with ISLiD components on one combined optical
bench in the central compartment of the PFI structure. Details of the SuFI
instrument, which has been developed by MPS, are given by Gandorfer \etal,
2010.

\subsubsection{Imaging Magnetograph eXperiment: IMaX}

IMaX is an imaging vector magnetograph for observations of Doppler shifts and
polarization in the Zeeman-sensitive photospheric spectral line of neutral
iron at 525.02~nm. Images are taken in up to twelve wavelength bands. The
instrument provides fast-cadence, high-spatial resolution two-dimensional
maps of the magnetic vector, the line-of-sight velocity, and continuum
intensity. It has a field-of-view of 50$\times$50~arcsec$^2$, which is the
largest of all instruments on \Sunrise. A tunable LiNbO$_3$ solid state
Fabry-P\'erot etalon is used in double pass, thus minimizing mass and power
consumption and relaxing the requirements on passband stability. Since the
free spectral range of such a system is quite small, a narrowband
interference filter (FWHM 0.1~nm) is used in addition. Both, prefilter and
etalon are thermally stabilized. Imaging is done with two synchronized
1k$\times$1k CCD cameras in orthogonal polarization states. Two nematic
liquid crystal retarders modulate the incoming polarization. Four switching
states are needed for full Stokes vector polarimetry, while two-states are
sufficient for longitudinal magnetometry (circular polarization). IMaX was
developed by a Spanish consortium led by the Instituto de Astrof\'{i}sica de
Canarias, La Laguna (Tenerife), in cooperation with the Instituto de
Astrof\'{i}sica de Andaluc\'{i}a, Granada, the Instituto Nacional de Tecnicas
Aeroespaciales, Madrid, and the Grupo de Astronomia y Ciencias del Espacio,
Valencia. The instrument is described by Mart\'{i}nez Pillet \etal, 2010.

\begin{figure}[t]
 \centerline{\includegraphics[width=1.\textwidth,clip=]{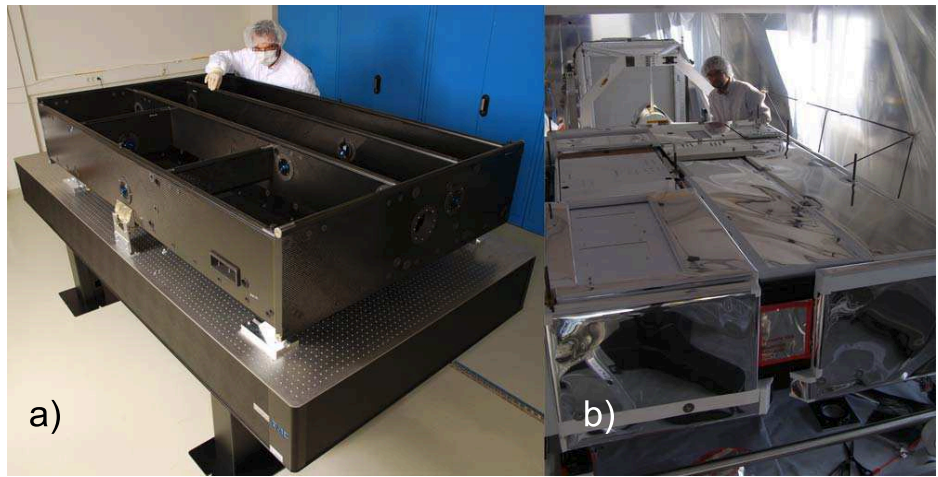}}
\caption{
a) Integrated PFI Support Structure without instrumentation mounted on optical bench, attached at all 7 interface points.
b) Fully assembled PFI in thermal insulation ready to be integrated onto telescope. }
\label{PFI_photo}
 \end{figure}

\subsubsection{PFI Support Structure}

The PFI structure shown in Figure~\ref{PFI_photo} provides a stable platform
for the science instrumentation. The individual instruments must be kept
co-aligned to better than $\pm$0.1~mm. This is a demanding requirement
considering instrument masses close to 50~kg (in case of IMaX) and in view of
the varying gravity and thermal loads during the mission. In addition, the
combined PFI package has to maintain the alignment with respect to the
telescope exit beam entering the PFI via M4. Similar to the telescope, carbon
fiber based composites and honeycomb structures are used for the mechanical
frame. This choice provides high stiffness and low thermal expansion,
together with low weight. The PFI structure has a width of approximately
1.4~m (compatible with the width of the telescope) and a length of about 2~m,
suitable to accommodate the instruments. The height of 350~mm resulted from a
trade-off between keeping the PFI center-of-gravity (CoG) as low as possible
and taking advantage of the increasing bending and torsional stiffness with
height.

Four outer panels form a box-like structure. Two inner panels spanning the
full length of the platform provide additional support for the inner
instruments and for the plane folding mirror M4, which is attached to the
rear PFI panel. Each panel is a combination of a 30~mm thick aluminum
honeycomb core with carbon-fiber-reinforced plastic face sheets of 1.25~mm
thickness. Dimensional stability of the panels is provided by a
($\pm$45$^\circ$, 0$^\circ$/90$^\circ$, $\pm$45$^\circ$) orientation of the
individual Torayca M40J fiber-layers of each facesheet. The coefficient for
thermal expansion (CTE) of the material is on the order of
$-$0.3$\times$10$^{-6}$ per Kelvin. To increase bending stiffness each panel
is reinforced at the two long sides by tubes with unidirectional fiber
orientation. Tenax UMS2526 fibers with an exceptionally high tensile strength
of 190~GPa are used. The tubes have 30~mm diameter and 1.5~mm wall thickness
each.

Additional cross panels stiffen out the compartment for the heavy IMaX
instrument. Shear stability of the box is provided by two panels of 10~mm
thickness, closing the bottom end towards the telescope. A central carbon
fiber based stiffener forms the central mounting point to the telescope
central frame. All connections are realized by steel screws and 2-parts
aluminum inserts, which are inserted into the sandwich panels from both sides
and secured by glue. The total weight of the PFI structure is 68~kg including
the mounting brackets for the instrumentation.

The center of gravity of the PFI is located directly above the steel
telescope central frame (Figure~\ref{telescope_CAD}c). A spherical pin at the
upper center of the telescope central frame is used to fix all three
translational degrees of freedom (DoF). Two flexural joints at the outer
uppermost ends of the telescope central frame fix the rotational DoF around
the Y-axis and Z-axis (direction to Sun), but allowing differential thermal
expansions of PFI and telescope structural elements in X-direction, caused by
the CTE mismatch of steel and the carbon fiber matrix.

Carbon fiber based struts support all four edges of the PFI, fixing the
rotational DoF around the X-axis (elevation axis). These struts have length
compensators eliminating any bending forces on the PFI structure in case of
temperature inhomogeneities between the telescope central frame and PFI (see
Figure~\ref{telescope_CAD}).

\subsubsection{Instrument Mounting}

All instrument units are individually assembled, aligned and functionally as
well as environmentally tested as separate modules before being integrated
into the PFI support structure. With the enormous effective focal length of
SuFI of 121~m even a minimal mechanical deformation on the order of only a
few micrometers already would transform into a considerable image shift on
the SuFI detector. The required stability cannot be provided by the 5~cm
thick optical bench alone. Increasing the overall stability of this sensitive
arrangement, ISLiD/SUFI makes use of the high stiffness of the PFI support
structure by being rigidly connected to the innermost side panels over its
full length. L-shaped brackets connected to the lower side of the ISLiD/SuFI
optical bench (Figure~\ref{ISLID_path}) provide a stiff mounting, but
allowing for local deformations in the vicinity of the attachment points, not
propagating to the surface of the optical bench or to the optical elements
fixed to it. Differential thermo-mechanical expansions of the ISLiD/SuFI
optical bench relative to the PFI platform are minimized by using face sheet
material from the same lot for both applications.

IMaX and CWS are mounted based on an isostatic concept, allowing the
individual alignment of each instrument module as such with respect to the
exiting beam coming from ISLiD. This concept has the advantage that optical
tolerances in focus position and exit beam orientation, as well as
unavoidable manufacturing tolerances of the large PFI support structure can
be easily compensated. The isostatic mounting provides a stress-free fixation
of the instruments even in the case of temperature differences and materials
with different CTE's. This is especially important for IMaX, which uses an
aluminum optical bench.

The correct angular orientation of IMaX and CWS with respect to the exiting
beams of ISLiD is achieved by the use of reflective alignment reference
cubes. Dedicated observation openings in the PFI structure panels allow
theodolite autocollimation measurements to arcsec precision.

A 3D coordinate measurement device (Leica Laser Tracker) led to a mounting
accuracy of better than 0.1~mm.

\subsection{Gondola}

The \Sunrise gondola is responsible for the precision pointing of the
telescope towards the Sun. The mechanical structure serves as stable platform
and protects the scientific instrumentation during launch and landing.
Photovoltaic arrays generate the required electrical power.

\subsubsection{Gondola structure}

\begin{figure}[h]
 \centerline{\includegraphics[width=1.\textwidth,clip=]{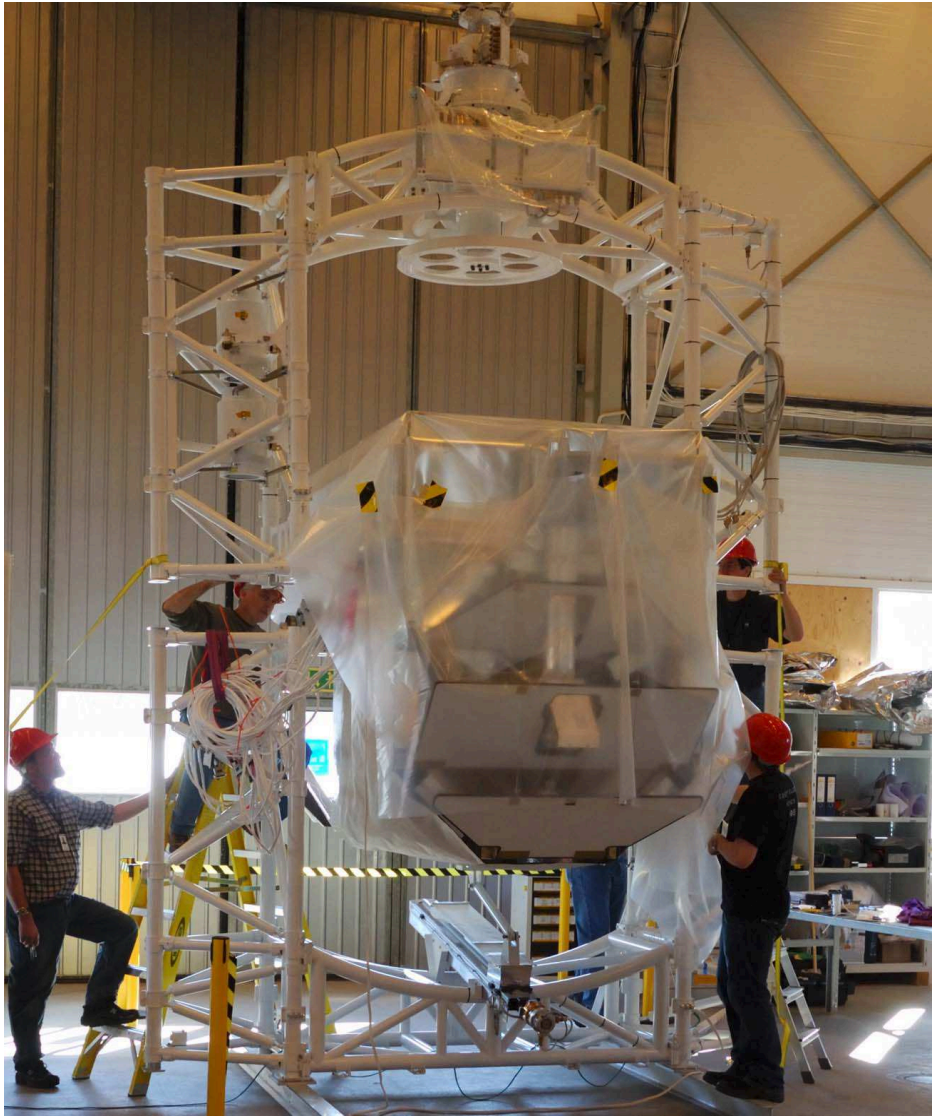}}
\caption{\Sunrise gondola main structure during integration of the science
payload. The two cylinders in the upper left side truss contain the science
data storage systems.} \label{gondola_1}
 \end{figure}

The core gondola structure consists of an aluminum/steel tube framework
(Figure~\ref{overview}), making it relatively lightweight, but providing the
required stiffness and a sufficiently high eigenfrequency ($>$10~Hz). The
structure permits a telescope elevation range of -5$^\circ$ to +50$^\circ$,
leaving some margin for pendulum motion during flight. The structure can be
split in two halves at the level of the telescope elevation axis
(Figure~\ref{gondola_1}), resulting in two U-shaped components. This allows a
convenient integration of the science instrumentation and simplifies shipment
of the bulky equipment. Each of these two units consists of two aluminum side
trusses with triangular cross-section; a welded steel-based bridge connects
the two. Front and rear roll cages are attached to the framework structure.
They protect the protruding front and rear ends of the telescope
(Figure~\ref{overview}). Below the bottom of the framework, the crash pad
assembly forms the mounting base for the cardboard landing-shock absorbers.
This frame houses the CSBF provided SIP (System Instrumentation Package), the
communication electronics needed for commanding to the instrument and
balloon, as well as telemetry through TDRS and Iridium satellite links.

\subsubsection{Power system}

Electrical power is provided by photovoltaic arrays on the front side of the
gondola, sufficiently far away that the nearly 100$^\circ$C hot panel
surfaces do not generate seeing effects (Figure~\ref{overview}). An
inclination angle of 22.5$^\circ$ optimizes the orientation for the expected
solar elevation range (0$^\circ$ to 45$^\circ$) during the flight. Each array
consists of five subframes, each with 80 A-300 cells, produced by Sunpower
Corporation. The panels, assembled by Meer Instruments, San Diego, USA,
generate approximately 1.3~kW in total. Two lithium-ion battery packs with a
capacity of 2500~Wh each act as buffers for the operation of the observatory,
before the correct pointing is achieved. The battery packs are mounted into
the lower bridge of the core framework.

\subsubsection{Pointing system}

\begin{figure}[h]
 \centerline{\includegraphics[width=1.\textwidth,clip=]{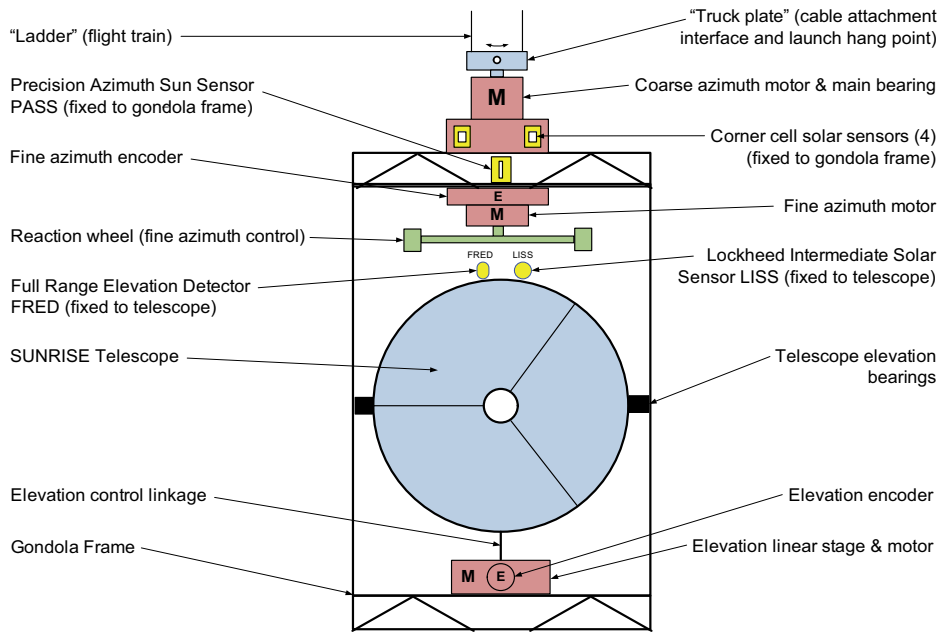}}
\caption{Basic \Sunrise gondola components}
\label{gondola_2}
 \end{figure}

The major components of the gondola pointing system are schematically
displayed in Figure~\ref{gondola_2}. Actuators and their encoders are shown
in pale red, solar sensors in yellow.

{\em Actuators:} Azimuthal control of the gondola is performed via a
two-stage momentum transfer unit at the top. The coarse azimuth drive
decouples the instrument from the rotating balloon, while the fine drive
rotates a reaction wheel. Acceleration or deceleration of the wheel rotates
the gondola towards the desired orientation. Nominal rotation rate of the
reaction wheel is 10~rpm, helping to overcome friction effects in the wheel
bearing. The actuation between the fine azimuth motor (the reaction wheel)
and the coarse azimuth motor is divided by frequency. The coarse azimuth
motor manages the frequency range below 0.1~Hz. The reaction wheel
compensates torques from $\sim$0.1~to $\sim$1.0~Hz. While the coarse azimuth
motor has a very large compensation range (it can apply about 13.5~Nm of
torque for an indefinite period of time), the fine azimuth motor can apply an
equal torque to the payload only for a limited period. Tracking is kept and
the fine azimuth servo system does not saturate, as long as disturbances to
the gondola -- like wind gusts -- induce less or equal torque for periods of
less than $\sim$1~second.

The elevation of the telescope is adjusted by a linear translation stage,
rigidly coupled to the central frame of the telescope by a lever arm. Slight
preloading of the lever arm minimizes hysteresis for the elevation
adjustment. Several elevation drive mechanisms were considered in the early
design phase, including torque motors mounted to the elevation axle. The
linear stage was chosen because it provides high precision and a relatively
non-compliant, symmetric loading on the central frame of the telescope. The
disadvantage of this system is additional weight over a direct drive torque
motor. However, the chosen solution serves as a launch and landing lock for
the telescope in (horizontal) stow position, without the need of an
additional mechanism.

Three types of sensors with different accuracy levels feed the various
control loops used to acquire correct solar pointing.

{\em Coarse gondola orientation:} Photovoltaic cells with wide-field,
overlapping acceptance cones of 60$^\circ$ in elevation and 105$^\circ$ in
azimuth are fixed to the gondola structure top at four corners. The right
quadrant is derived from comparison of intensity levels, independent of solar
elevation.

{\em Medium azimuth \& elevation detection:} The Precision Azimuth Sun Sensor
'PASS' is also fixed to the gondola structure top. The sensor is used as
medium resolution device, locating the Sun in azimuth with a linear capture
range of $\pm$3$^\circ$ for any given solar elevation. For medium resolution
detection of the Sun's elevation, a similar device is used. The Full Range
Elevation Detector 'FRED', mounted to the telecope front ring, can find the
Sun within an azimuth range of $\pm$5$^\circ$. This elevation sensor has a
linear range of $\pm$15$^\circ$. For angles outside this range, it gives a
saturated signal indicating that the Sun is above or below the current
telescope elevation.

{\em High resolution solar position sensing:} Highest accuracy in determining
the solar position is provided by the LISS (Lockheed Intermediate Sun
Sensor), mounted to the \Sunrise telescope front ring. The LISS is used
within a $\pm$3$^\circ$ range in azimuth and elevation, where the output
signal is proportional to the off-pointing from the LISS optical axis. The
high resolution linear range is only $\pm$15~arcmin. Electronic noise and
signal drifts with temperature are low enough, so that accuracies of
1-2~arcsec can be achieved. The LISS is mounted on a two axis motorized
tip-/tilt stage. Offsets in telescope pointing for solar limb observations
and instrument flatfielding are generated by commanding position offsets to
the LISS motors.

The medium and high resolution sensors PASS, FRED and LISS are shadow sensor
type detectors. In case of the LISS, solar radiation passes a square aperture
entrance window, falling on a set of 5~photodiodes. A central diode is used
as a binary 'Sun present' indicator. The remaining 4 sensors are arranged in
two pairs, oriented in azimuth and elevation direction. Each pair is
connected to a transconductance amplifier, which amplifies the difference in
photocurrent. When the sensor is tipped, the amount of light falling onto the
two photocells (in a given axis) gets out of balance and produces a signal.

{\em Acquisition of solar pointing:} The acquisition of solar pointing and
tracking is completely automatic. Several software threads handle data
collection, processing, actuator stimulation and data output generation. The
'auto-pointing thread' makes the decisions of which Sun sensors to use and
what servo control loops to engage. The 'interrupt thread' runs at 150~Hz
update rate and does all the real time pointing and data collection. It also
records a boxcar array of sums and sums squared of all Sun sensor information
and voltages for the elevation and coarse and fine azimuth motors. A separate
thread is constantly checking thresholds to determine wether the pointing
system is locked on the Sun and can set the 'PS-lock' flag for the CWS image
stabilization system (see Section 2.2.2.).

The auto-pointing thread uses the sums and sums squared to compute running
averages and standard deviations squared of these parameters. The running
averages and standard deviations are computed based on the most recent
1024~values collected, (1024~samples covers 6.8~seconds of data). The
auto-pointing thread checks every 9~seconds all instantaneous values, running
averages and running standard deviations. Then, using carefully adjusted
threshold criteria, it decides which acquisition sequence needs to be
activated.

The threshold criteria and activation of servos depend on which solar sensor
is getting the best signal. Each pointing sequence aims to optimize the
positioning for the next higher resolution sensor. The automated acquisition
uses the following sequential steps with dedicated control loops:

\begin{itemize}
\item Coarse azimuth: Coarse+fine azimuth drives to find correct azimuthal
orientation based on corner cell signals, if gondola rotation rate is
less than 2~rpm
\item Fine azimuth: Coarse and fine azimuth drives to position gondola based on
PASS data
\item Coarse elevation: Telescope elevation is adjusted according to FRED data,
until LISS sensor indicates Sun presence
\item Final tracking: Control loops closed in azimuth and elevation, based on LISS
data
\end{itemize}

\begin{figure}[h]
 \centerline{\includegraphics[width=1.\textwidth,clip=]{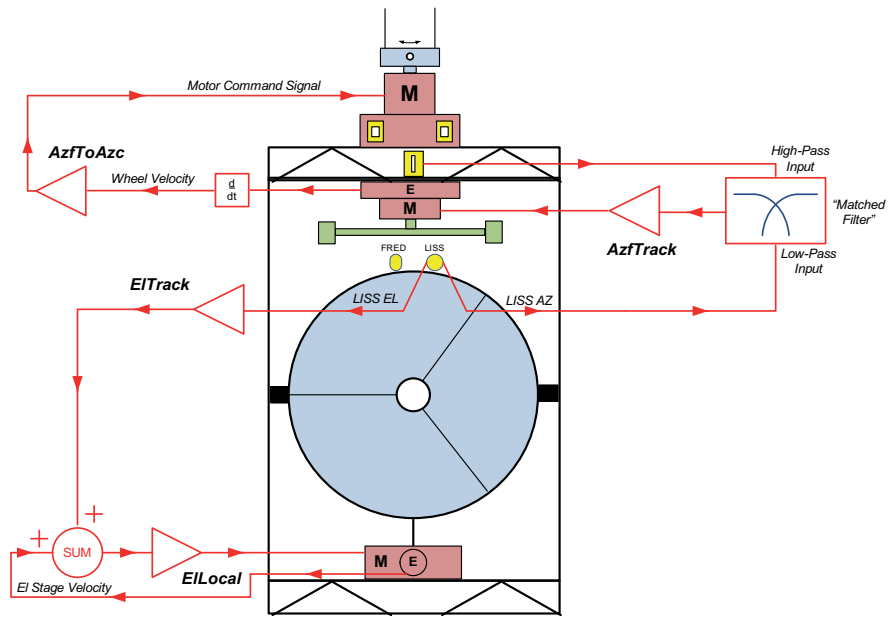}}
\caption{\Sunrise final tracking servo control loop configuration.
Red lines represent signal paths and triangles are signal processing (filtering).
Yellow elements are solar sensors. Motors and encoders are shown in pale red. The label 'AzfToAzc' denotes the servo loop monitoring the rotational velocity
of the fine azimuth drive and providing feedback to the coarse azimuth motor to eventually slow down the reaction wheel.
Similarly, 'AzfTrack' and 'ElTrack' denote fine tracking servo loops. 'ElLocal' transforms input values to calibrated velocity commands for the elevation drive.}
\label{gondola_3}
 \end{figure}

Final tracking is achieved by switching from 'FRED' to 'LISS' for elevation
tracking and from the 'PASS' to 'LISS' for azimuth tracking. This switching
in azimuth and elevation is done separately to accommodate which sensor gets
to a good signal first. In azimuth the pointing system uses a
'multiple-input-multiple-output' (MIMO) approach. The final servo control
loop topology is shown in Figure~\ref{gondola_3}. The filters settings
applied in this mode have a higher gain and are designed for a better
tracking accuracy. The servo loops use a combination of integrators and
phase-lead compensators with up to four filters per loop. In case of the
'AzfTrack' servo loop (see Figure~\ref{gondola_3}) a combination of a second
order low pass filter, then a first order lead filter and lastly a first
order integrator filter is applied to the LISS azimuth output, driving the
reaction wheel.

The design requirement for the tracking accuracy is $\pm$7.5~arcsec rms, to
safely keep the system within the capture range of the image stabilization
system, provided by the tip-tilt mirror and CWS ($\pm$46 arcsec).

The automated acquisition of the Sun from a non-pointing mode to 'pointing
system lock' is typically achieved within less than 10 ~minutes.

Matlab models were used to try to predict the behavior of the full flight
train versus the suspension used during ground testing. The main mitigation
was to not use positional feedback on the coarse azimuth drive. Such feedback
gives rigid coupling to the flight train and thus the vibrational modes of
the ladder are very well coupled to the gondola. Torque feedback was used
instead (simple current to the motor), assuming that the main rotator bearing
was good enough that ladder modes would be decoupled. Servo filter settings
derived from the mathematical models were uploaded during the commissioning
phase, being fine-tuned throughout the mission.

\subsubsection{Gondola mounted components/systems}

On the rear side of the gondola, shaded by the solar panels from direct Sun
illumination, the instrument control electronics are located on two racks.
One rack houses the instrument control unit ICU, the payload power
distribution unit and the instrument computers for CWS, IMaX and SuFI. The
other rack carries the pointing system computer, the gondola power
distribution unit, amplifiers for the azimuth and elevation drives, as well
as the ESRANGE-provided E-Link high-speed telemetry electronics. The racks
are inclined with respect to the structure in order to minimize radiative
input from the Earth and the hot solar panels onto the electronics, while
maximizing the radiation of heat to the cold sky above the instrument. The
two data storage containers collecting data from the instruments are mounted
well secured inside one of the upper side trusses of the core framework. The
truss framework provides protection, but also allows easy access for recovery
after landing. A spring-based shock protection system ensures the mechanical
integrity of the data storage containers. To avoid oscillating masses and
adverse effects on the pointing control loops, the data storage containers
are rigidly clamped during the mission. A release mechanism frees the
containers at mission termination.

The commanding and communication package provided by CSBF is located
underneath the gondola structure (Figure~\ref{overview}). It has separate
solar panels at all four sides of the gondola to stay operational in case of
pointing loss. The package allows commanding and housekeeping downlink via
TDRS and Iridium satellites. Shock absorbing cardboard crash pads at the
bottom of the gondola reduce mechanical loads during touch-down and landing.
Two ballast hoppers at the center of the gondola bottom carry about 650~kg of
fine steel grains. Part of the ballast is dropped during ascent, speeding up
the balloon again after it has cooled down in the tropopause transit. The
rest of the ballast is used to compensate losses in float altitude due to the
day/night cycles. Two booms at the top of the gondola carry the satellite
communication antennae.

The complete payload has dimensions of 5.5 meters in width and length and is
about 6.4 meters high. The gondola structure, the power and pointing systems
have been developed by the High Altitude Observatory (HAO), NCAR, Boulder,
USA.

\subsection{Electronics Architecture}

The largest part of the \Sunrise electronics is located on two racks mounted
left and right of the gondola structure. Only proximity electronics such as
mechanism controllers or the voltage supply for the piezo-driven tip-tilt
mirror are located close to the optical modules inside the PFI.
Commercial-off-the-shelf products are used as far as possible. These products
would typically not survive the environmental conditions of a balloon flight.
Critical items therefore were encapsulated in pressure vessels, modified or
specifically qualified for this type of application.

\Sunrise is designed for autonomous operation similar to a spacecraft. The
basic architecture consists of an instrument control system
communicating with dedicated subsystem- or instrument-related electronics,
and telemetry systems for commanding and downlink of system status
information.

The instrument control system consists of the Instrument Control Unit (ICU), two data storage
subsystems, the PFI and gondola power distribution units, and the line-of-sight telemetry subsystem (E-Link).

The ICU is the central onboard computer. It supervises the various subsystem
and instrument control computers:
\begin{itemize}
\item Pointing system computer
\item Gondola power distribution unit
\item CWS electronics unit
\item Main telescope controller
\item SuFI electronics unit
\item IMaX main electronics
\item Payload power distribution unit
\item E-Link
\item PFI mechanism controllers (also used for PFI thermal control)
\end{itemize}

The instrument control system carries out the following tasks:
\begin{itemize}
\item interpret and route telemetry commands to subsystem and instrument computers
\item acquire housekeeping data
\item store housekeeping and science image data on data storage units in RAID5 format
\item pre-select and route housekeeping and science image thumbnail data to downlink telemetry subsystem
\item initialize, configure and run the \Sunrise instrument automatically in standard observational mode, if  no observation mode is given by
telecommand
\item run predefined timeline-controlled observations
\item control electrical power distribution to scientific instruments and subsystems
\item monitor housekeeping data and take action in case of limit violations
\item provide line-of-sight telemetry
\end{itemize}

The Instrument Control Unit (ICU) is housed in a pressurized container. It is based on a NOVA 7800 P800
Pentium III single board computer with interface cards providing ethernet and
serial interfaces. A 2 GB flash disk serves as boot device. Forced convective
cooling by regulated fans keeps the component temperatures within specified
ranges. Onboard communication between ICU and subsystem and instrument
computers is performed via a 100base TX ethernet and serial RS232 resp. RS422
links.

The science data are stored onboard in two stacks with 24 (100~GByte)
harddisks each with about 3.6 Terabyte net capacity at RAID-5 functionality.
This corresponds to the expected data volume acquired in a 2 weeks mission.
The harddisks are encapsuled in two pressurized vessels (see Figure~\ref{gondola_1}), maintaining the
required environment regarding temperature, pressure and humidity. The data
from the ICU to the DSS are transferred via IEEE1394 A (Fire Wire) links at
400~Mbit/s.

The commanding and telemetry from and to the ground station is handled via
the CSBF-provided System Instrumentation Package (SIP). The SIP provides
simple pulse command channels for direct instrument control. The pulse
commands are used for ICU reset control and for payload power switch off in
case of emergency or to release S/W induced deadlocks.

\subsection{Telemetry}

Operations control for \Sunrise is performed at two different locations. All
balloon system relevant issues such as flight monitoring and tracking,
ballast drops and mission termination are handled by the Operations Control
Center (OCC) in Palestine, Texas, USA. \Sunrise science operations are
directly controlled and monitored from system and instrument EGSE (electronic
ground support equipment) computers at the Remote Operations Control Center
(ROCC) located at ESRANGE. All command activities at ESRANGE are closely
coordinated with CSBF personel present on site.

Commanding and downlink of telemetry data are provided through different
communication systems depending on the mission phase.

During the first hours after launch, while the instrument still is within the
line-of-sight, the E-Link high speed telemetry system is used. This
communication system has been developed by ESRANGE and is available on a
rental basis. It operates at frequencies around 2.4~GHz and acts as a
transparent Ethernet connection to the ICU, providing simultaneous up- and
downlink rates of up to 2 Mbit/s. The range is limited to approximately
350~km. The ground station antenna tracks the instrument position
autonomously using the GPS information provided onboard \Sunrise. To enhance
the range for line-of-sight communication, a second ground station was set up in
Andenes, Norway.

\begin{figure}[t]
 \centerline{\includegraphics[width=1.\textwidth,clip=]{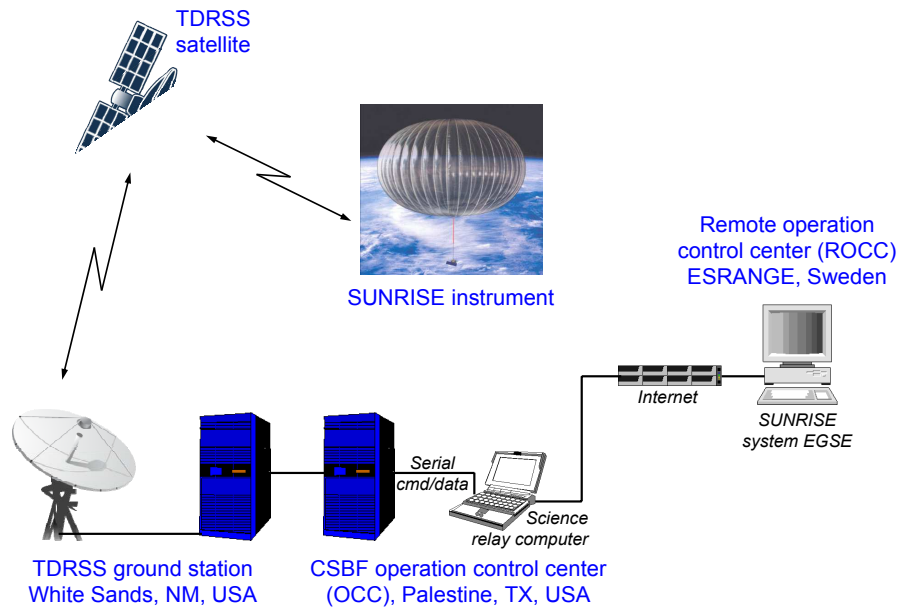}}
\caption{'Over the horizon' communication via TDRSS satellite link}
\label{TDRSS}
 \end{figure}

`Over the horizon' communication with \Sunrise is achieved by a Tracking and
Data Relay Satellite System (TDRSS) communication channel. Commands generated
by the \Sunrise EGSE computers at ESRANGE are transferred via internet to a
science relay computer at the OCC in Palestine, Texas ( Figure~\ref{TDRSS}).
The OCC is connected with the TDRSS ground station in White Sands, New
Mexico, where the \Sunrise commands are uplinked to the TDRS and finally
relayed to the SIP antennas onboard Sunrise. The SIP transfers the commands
via a serial link to the ICU. Data generated onboard, such as instrument
housekeeping or thumbnail images, are relayed from the instrument back to the
ROCC along the same path. A data rate of 6~kbit/s is available almost
permanently for the downlink.

\subsection{Balloon System}

\begin{figure}[t]
 \centerline{\includegraphics[width=1.\textwidth,clip=]{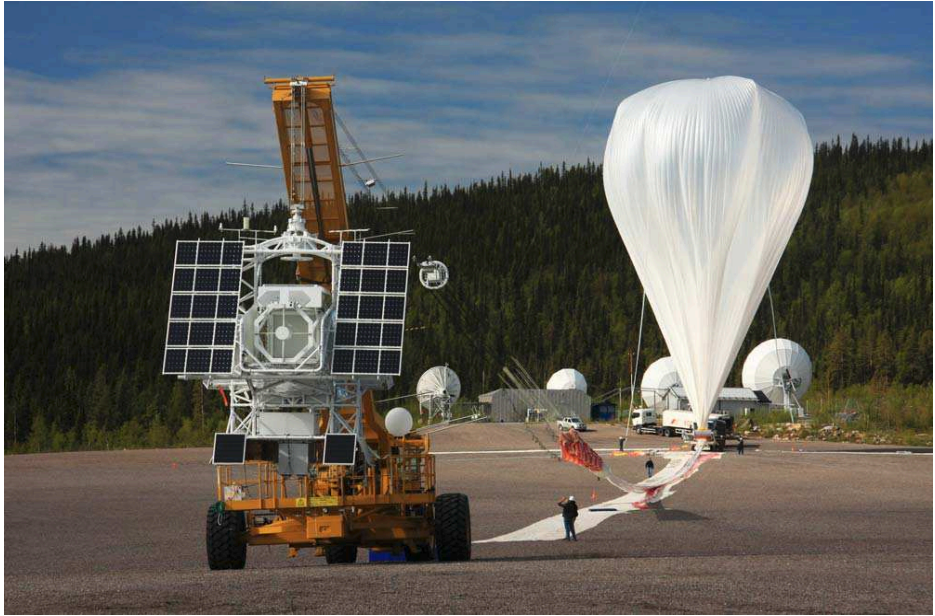}}
\caption{Sunrise immediately before launch. The balloon is only partly filled on ground to allow expansion  with decreasing outside air
pressure during ascent. At 37~km altitude it will have expanded by a factor of 300 as compared to ground. The parachute for landing is part of the flight train connecting the gondola payload with the balloon.}
\label{balloon}
 \end{figure}

The science goals of \Sunrise profit from as high a float altitude as
possible, especially to allow observations in the UV down to 214~nm.
Therefore one of the largest balloons regularly flown by CSBF has been chosen
for \Sunrise, an Aerostar zero-pressure balloon with a volume of
34.43~million cubic feet (975.000~cubic meters) and a diameter of close to
134~meters. In case of \Sunrise this balloon type lifts a total mass of about
6 tons to stratospheric altitudes of more than 37~km. The scientific payload
with a mass of 1920 kg contributes only one third to the total weight. The
rest is given by the balloon film (2330~kg) with its helium filling
(~500~kg), auxiliary equipment as suspension, parachute, crush pads, ballast
hoppers (Figure~\ref{balloon}) and finally ballast (544~kg) for altitude
stabilization during flight.

\subsection{System Analysis and Design Aspects}

The optical performance of \Sunrise must be guaranteed under various thermal
and mechanical load cases. In order to stay within the required optical
tolerances, which were defined by a sensitivity and tolerance analysis on
telescope and instrument level,
  the \Sunrise design was based on a thorough analysis of the
expected thermal and mechanical loadcases.

\subsubsection{Structural Design}

A structural analysis and iterative optimization of the complete instrument
with all components of the gondola, telescope and PFI has been performed with
the aim of maximizing stability and stiffness, while minimizing weight. The
NASTRAN finite element software package has been used. Several requirements
had to be considered, such as instrument safety, alignment and pointing
stability:

- Safety aspects are important when bringing a payload of several tons to
stratospheric heights. CSBF requests to analytically prove instrument
integrity with respect to mechanical loads induced at launch and flight
termination. All components therefore have sufficient margin-of-safety to
structural failure, for instance regarding vertical shock loads of up to
10~$g$, expected during parachute opening.

- Mechanical loads induced during transport, ground handling and launch must
not lead to permanent misalignment of structural components or instrument
units. The design of all components and fixation devices considers a
quasistatic load of 4~$g$, leading to elastic deformations only.

- In contrast to a space instrument, \Sunrise is subjected to gravity when in
operation. The orientation of the telescope and instrumentation relative to
the gravity vector varies significantly with the changing elevation of the
Sun. This could affect the relative alignment of structural components or
instruments. The telescope Serrurier structure as well as the PFI support
structure provide sufficient structural stiffness to maintain optimal
instrument performance. A few residual thermo-elastic deformations cannot be
controlled purely by passive structural design. Therefore, the critical
relative positioning of telescope primary and secondary mirror with a
sensitivity in the micro-meter range is actively adjusted inflight, ensuring
the high performance image quality of the system.

- The servo control loops providing the gondola pointing assume rigid body
motions when activating the azimuth and elevation drives. Rigid body motion
is only given when the lowest eigenfrequency of the moving parts is
considerably higher than the excitation frequencies. As the elevation and
azimuth drives operate at frequencies below 10~Hz, the gondola as largest
structure has been designed to a minimum eigenfrequency of the order of
15~Hz. Decoupling of eigenfrequencies avoids excitation of one sub-system by
another. The PFI has been designed to a minimum eigenfrequency of 25~Hz and
the telescope to 35~Hz.

\subsubsection{Thermal Design}

Thermal analysis and thermal design is important to keep all instrument
components within their specified temperature ranges, under all conditions.
This is especially relevant for the high power dissipating commercial
electronics and alignment-critical optics. The environmental conditions for
Sunrise at stratospheric altitudes are very different compared to ground.
Owing to the low pressure of only a few millibars, convective coupling is no
longer the dominant heat exchange mechanism and the thermal behaviour of the
instrument is controlled practically only by radiative heat exchange.
Component temperatures result from the equilibrium of dissipated, absorbed,
and emitted energy. Surface properties such as absorption and emission
coefficients as well as view factors of surface elements with respect to
their neighbourhood and to heat sources and sinks are dominant factors.

During flight, \Sunrise is exposed to a changing environment. The albedo and
thus the heat input from below significantly varies when flying over sea, ice
or cloud layers. In addition, the radiative input varies with solar elevation
angle.

Two extreme steady state load cases covering the wide range of conditions to
be expected were defined:
\begin{itemize}
\item a 'hot' case with 45 deg solar elevation angle, a solar flux of 1397~W/m$^2$, an albedo of 0.95 simulating complete ice
coverage, and Earth IR radiative input of 264~W/m$^2$
\item a 'cold' case with 0 deg solar elevation angle, a reduced solar flux of 1044~W/m$^2$, an albedo of 0.11 simulating sea water below
the instrument, and Earth radiative input of 156~W/m$^2$.
\end{itemize}
\Sunrise with all components was modelled with several thousand nodes using
the ESA\-RAD/ESA\-TAN software package. Geometries, surface properties and
radiator sizes were determined and optimized.

{\em Thermal Control of Instrument Electronics and Structural Components:}
Heat dissipating elements such as computers, photovoltaic arrays etc. have
been placed as far away as possible from the telescope and instruments. The
instrument electronics are located on two aluminum honeycomb racks of
1$\times$2m$^2$ size, left and right on the rear side of the gondola. The
racks are shaded by, and turned away from the solar arrays, minimizing the
view factor of the electronics to the nearly 100$^\circ$C hot photovoltaic
cells. A 20$^\circ$ inclination reduces albedo input and maximizes the view
to the cold sky, while still keeping the moments of inertia of the system
low. White paint is used as surface treatment for all the electronics units
and passive structural components directly exposed to the Sun, such as the
tubing of the gondola and the telescope mechanical components. Thermal filler
is used underneath some of the electronics, enhancing the conductive coupling
to the top surfaces of the racks so that they serve as additional radiating
areas.

{\em Gondola Blankets:} The impact of the changing environment on telescope
and instrumentation must be minimized to reduce thermo-mechanical
deformations. Shading solar radiation coming from the extended source below
the observatory is provided by large thermal blankets, which are mounted to
the gondola interior sides and bottom, including front and rear roll cage.
The outer blanket layer is facing the Earth and reduces the heat input to the
system. A Mylar foil of 125~$\mu$m (5 mil) thickness with vapour deposited
aluminum on one side was chosen. The foil is used as second surface mirror,
providing similar thermal properties as white paint. The innermost layers of
the gondola thermal blankets control the heat exchange with the telescope and
science instrumentation. They are directly illuminated by the Sun during
observation. Dunmore Beta Cloth 500F is used in the front roll cage. This
low-outgassing, space approved material is tailored to form a diffusely
reflecting white cavity around the front end of the telescope, reducing the
probability of structural hot spots. The interior of the rear roll cage is
covered with a highly reflective aluminized polyimide (Kapton) foil,
enhancing the radiative exchange of the primary mirror with the cold sky.
Here the aluminum side is facing outwards, maximizing the reflection
properties. Heating of the foil is only moderate, due to the grazing solar
incidence on the roll cage sides and the shadowing by the telescope and
instrumentation.

{\em PFI Thermal Design:} The thermal design of the postfocus instrumentation
places emphasis on thermal stability. The overall temperature level including
gradients across the 2 meter long structure have to be kept within a small
temperature range of 20$\pm$10$^\circ$C. This ensures not only invariant
alignment for non-carbon-fiber-based instruments as IMaX, but also minimizes changes
of the polarization properties of the optics.

The PFI structure is decoupled from the environment by Styrofoam insulation,
a low outgassing closed-cell material. Panels of 4~cm thickness are used on
all 4 sides of the PFI, and 1~cm panels on the bottom towards the telescope.
The panels are tailored to fit to the structure and wrapped and taped in
aluminized Mylar foil of 125~$\mu$m thickness, again used as second surface
mirror. The foil wrapping minimizes particulate contamination of the optics,
which could occur when using cut foam pieces. The top side of the PFI has a
large view factor to the heat sink of the cold sky above the instrument.
Structural components are covered with wrapped Styrofoam, similar to the PFI
sides and bottom. White painted radiator plates serve as direct link for
heat-dissipating proximity electronics, such as mechanism controllers, drive
electronics for the tip-tilt mirror and SuFI camera head, electronics and
power supply (Figure~\ref{PFI_photo}b). Fine adjustment of the expected
electronics temperatures was achieved by partially covering the radiator
areas with wrapped foam pieces. Groups of thermostat-controlled and actively
controlled foil heaters were placed on the metallic covers of the ISLiD,
SuFI, and CWS instruments underneath the foam insulation. A heat shield on
the illuminated front end shades the PFI from direct solar radiation and
minimizes temperature gradients in addition to the foam insulation.

{\em Wind Shields:} The balloon ascent through the cold tropopause layer is
considered as a risk for the PFI instrumentation and instrument electronics.
Convective cooling in conjunction with the radiative heat exchange can lead
to a critical temperature drop during this phase (see below). Dedicated wind
shields were developed to reduce the impact of the cold air stream while not
adversely affecting the thermal conditions at float. A 37.5~$\mu$m (1.5 mil)
Polyethylene film from the NASA ULDB program was selected and kindly provided
by Aerostar Industries free of charge. This material is optimized in terms of
mechanical durability at low temperature conditions and provides a high
transparency in the visible and the infrared. The wind shield material and
mounting concept was tested to withstand wind gusts and velocities up to
100~km/h.

\section{Mission}

\subsection{Subsystem Test Flights}

The gondola system and large parts of the instrument control system including
the main computer (ICU), data storages and power distribution units,
performed a 9-hour test flight on Oct 3, 2007 from Ft. Sumner, New Mexico,
USA. The gondola was equipped with a small UV telescope with 25~cm aperture
and a camera system with a filter wheel, verifying the pointing capabilities
of the gondola. Fully powered by batteries, the gondola carried solar panel
dummies, simulating the thermal behaviour and being representative in terms
of size, mass, inertia and aerodynamics. Pointing, onboard data acquisition,
commanding and downlink through telemetry systems were tested. The payload
was recovered largely intact on a wheat field close to Dalhart, Texas, USA,
the following day.

The test flight revealed the sensitivity of some of the electronic equipment
when exposed to the severe environmental conditions around the tropopause.
Low temperatures of less than -75$^\circ$C (Perez Grande \etal, 2007) caused
the ICU to temporarily stop working. The ICU recovered at float altitude and
performed flawlessly during nominal flight operation. For the science flight
it was decided to modify the thermal design of the Sunrise instrument by
implementing dedicated wind shields (see Section 2.7.2). A verification of
this concept was performed on a dedicated stratospheric balloon test flight
of electronics equipment in June 2008 from ESRANGE.

\subsection{Instrument Integration and Mission Preparations}

The two major components of \Sunrise, the gondola and the telescope with its
instrumentation package, have been assembled, integrated and tested
separately at their respective home institutions, before being shipped to
ESRANGE for system integration.

The gondola had survived the 2007 Ft. Sumner test flight relatively unharmed.
The refurbishment was done at HAO. A few mechanical components needed
replacement (crash pad assembly and roll cages) and some electrical
modifications were implemented at HAO. Significant work was spent improving
and testing the pointing control software.

At MPS the postfocus instrumentation (PFI) including ISLiD/SuFI was assembled
and tested. The two externally provided, pre-aligned instrument units CWS and
IMaX then were integrated and co-aligned relative to ISLiD/SuFI. Final verification on
PFI level included checks of the overall wavefront quality. Polarization
properties of the integrated package as well as the closed-loop performance
of the tip-/tilt mirror were measured.

The telescope was assembled, aligned and tested at Kayser-Threde, Munich,
under contract by MPS. After delivery to MPS end of 2008, all relevant
functional tests were repeated, including wavefront measurements with an
interferometric end-to-end test and a verification of M2 movements to
micrometer accuracy. The tests proved the excellent stability of the
telescope system and verified the transportation concept with a dedicated
damped telescope dolly tailored to fit into a standard sea container.

Telescope and postfocus instrumentation were mated for final instrument
performance tests at MPS. Lasertracker measurements helped to align the PFI
package on top of the central telescope frame to its nominal position. The
four struts supporting the PFI edges needed careful adjustment to avoid any
distortion or bending of the large structure, in order to reproduce the
alignment status achieved earlier on the optical table.

Due to transport size limitations the PFI had to be dismounted from the
telescope again for shipment. Both units were packed separately in vibration
damped transport dollies and stowed in two 20ft sea containers. All equipment
from HAO and MPS arrived end of March 2009 at the launch site ESRANGE.

The \Sunrise flight hardware and auxiliary equipment were set up in the
integration hangar nicknamed 'Cathedral' with approx. 250~m$^2$ floor space.
Functional tests of all units, alignment verification of the PFI and
interferometric wavefront measurements on the telescope proved that all
systems had survived the transport without degradations. PFI and telescope
were mated on April 11, 2009, integration into the gondola was done one week
later on April 18, 2009.

Extensive in-door testing was performed with the nearly fully assembled
instrument, being suspended by the hangar crane. Parameters of the pointing
system control loops were adjusted to the real moments of inertia of the
flight hardware. A 10 kW tungsten theater light was used as artificial sun.
Although not giving the radiance levels comparable to direct sunlight, the
response of all pointing relevant sensors such as corner cells, mid level
sensors and the fine guider could be tested and the system response was
optimized. Pointing tests were continued with real sunlight from within the
'Cathedral' the following days, further refining the control loop
performance, now with higher photon flux on the sensors.

\Sunrise instrument ''first light'' was achieved on April 30, 2009. Tests
verified the performance of the heat rejection wedge cooling, then first
images were taken with the science instruments. Due to the very poor ground
seeing conditions and the absence of sunspots no image stabilization was
possible as no features were visible on the Sun. However, the exposure times
for SuFI, IMaX and CWS could partly be verified in flatfield images, and the
important wavelength calibration of IMaX could be performed.

Autonomous operation of the instruments needs predefined timelines of
commanding. Those were extensively tested in May 2009, harmonizing and
optimizing the various software codes of the different instrument units.

A full flight configuration compatibility test was successfully conducted on
May 30, 2009 together with CSBF and ESRANGE. \Sunrise was moved from the
'Cathedral' to the outside, all additional flight equipment was mounted. The
test demonstrated electro-magnetic compatibility of all components and as
well mechanical compatibility with regard to the launch procedures. \Sunrise
declared flight readiness on June 5, 2009, after having performed a last sun
pointing test with IMaX calibrations.

\subsection{Flight June 2009}

\Sunrise was successfully launched on June 8, 2009 at 6:27~UT (08:27h local
time), on its first launch opportunity. Balloon and instrument reached an
initial float altitude of 37.2~km (122.000~ft) after about three hours
ascent.

Instrument commissioning started with the gondola pointing system. During
ascent, \Sunrise was in a rotary mode, providing solar illumination on
virtually all parts of the instrument and thus some heating during the
critical tropopause transit. This mode however required the whole system to
be run on the limited battery power. Having adjusted relevant parameters of
the pointing control loops to the different conditions due to the longer
flight train, the telescope aperture door was opened at 11:41:15~UT.
Instrument check-out and commissioning proceeded with the CWS, closing the
loop for image stabilization and adjusting the focus of the telescope first
at 13:20~UT. At that time, first images were taken by SuFI and IMaX. A full
commissioning of the instruments was, however, impossible due to an early
failure of the line-of-sight communication. The 2 Mbit/s high speed telemetry
link operated until 15:09~UT, but was unreliable towards the end. The
hand-over to the second ground station in Andenes/Norway unfortunately
failed. This additional ground station would have extended the communication
range to approximately 24 hours coverage, given the low speed and direction
\Sunrise was heading. The TDRSS link had to be used instead, providing a
downlink data rate of approximately 6 kbit/s. With this low data rate, some
of the originally foreseen commissioning tasks could not be performed as
planned.

The observation program collected minimum science data at disk center on June
08 and 09, 2009. On June 10 beginning at 01:36~UT co-alignment of telescope
and sun sensor LISS was checked by searching the solar limb. Instrument
flatfielding was performed by commanding a circular movement of the telescope
pointing with fixed tip-/tilt mirror.

\begin{figure}[t]
 \centerline{\includegraphics[width=1.\textwidth,clip=]{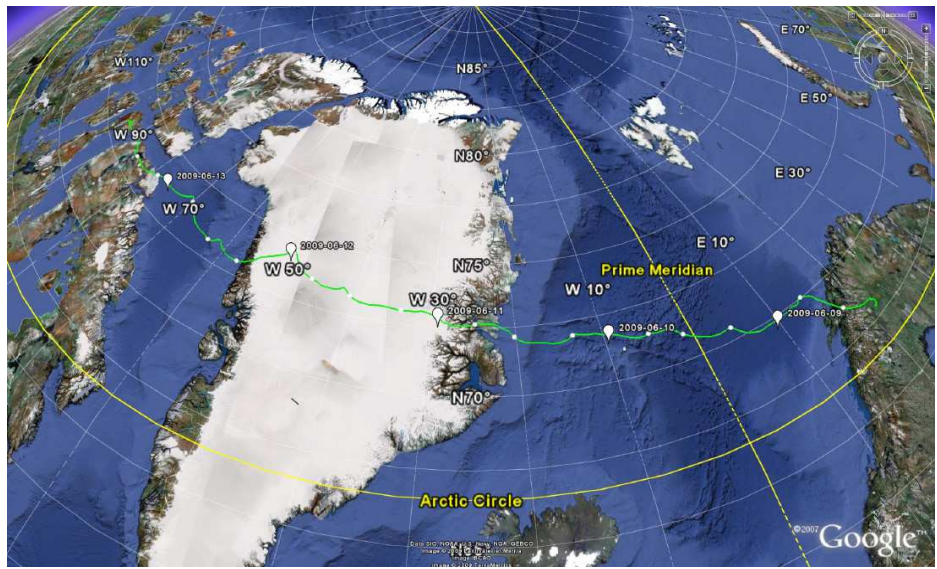}}
\caption{Flight trajectory of Sunrise.}
\label{trajectory_1}
 \end{figure}

Zonal winds carried the balloon and instrument with almost constant speed of
30~km per hour to Northern Canada (Figure~\ref{trajectory_1}). The balloon
altitude varied between 37~km and 34~km, following the sun elevation with a
phase lag of approximately 3 to 4~hours. Several ballast drops in the second
half of the mission helped to regain float altitude.

Flight termination sequences were commanded on June~13, 2009 at 21:20~UT ,
followed by the balloon cut-away at 22:52~UT. \Sunrise landed on Somerset
Island, Nunavut County, Northern Canada at 23:44~UT after 137 hours mission
elapsed time and a travel distance of nearly 4350~km.

\Sunrise was completely recovered a few days after landing. All instrument
parts were flown out via helicopter first to Resolute Bay, then with
airplanes to Yellowknife, where the equipment was packed into sea containers
for shipment to their home institutions. Damage to \Sunrise was found to be
moderate, \eg the primary mirror survived the landing perfectly intact. The
scientific data stored on the data storage harddrives safely arrived at MPS
on June~25, 2009, being handcarried directly from Yellowknife on commercial
airplanes.

\section{Instrument In-Flight Performance}

The \Sunrise instrument impressed with near flawless performance during its
maiden flight, although instrument commissioning could not be fully
accomplished due to the short period of line-of-sight high speed
communication (see previous Section).

{\em Power system, Electronics and Software:} All on-board electronics, as
power system, instrument computers, proximity electronics and their related
software, performed nominally during most of the mission.

The sizing of battery capacity and solar panels proved to be very
conservative. Although the time to first Sun acquisition was about two hours
longer than anticipated (see below), the battery charge state did not drop
below 77~\%. Once correctly oriented, the solar panels fully recharged the
batteries within 10 hours. The system current plus the charge current to the
batteries were provided by only 3-4 out of the 10 panels. The charge
controller automatically deactivated the unused panels.

Some issues were identified concerning the handling of the data storage
units. On the second day (June 9th) the instrument control unit had to be
rebooted. It was unable to recover from a failed write operation to one of
the active disks. During the data post processing, it became obvious that the
disk showed signs of malfunction already before that, since there were files
missing from the observations made during the first night. However, all data
were successfully recovered due to the applied RAID functionality.

{\em Thermal:} All component temperatures remained well within their
operational design limits and close to their predictions given by the
detailed thermal models.

During ascent all components covered by windshields --~PFI and electronics
racks~-- encountered only minimal temperature drops to about 0$^\circ$C, when
passing the cold tropopause layer. Temperatures of unshielded components, as
the gondola structure, the solar panels or the telescope trusses, dropped to
about -40$^\circ$C at the same time. At float, the temperatures for
instrument computers mounted to the electronics racks stayed within
0$^\circ$C and +30$^\circ$C, see Perez-Grande \etal,~2010 and Berkefeld
\etal,~2010. Similar temperatures were measured for the proximity
electronics, as mechanism controllers within the PFI structure. The optical components and
mechanisms inside the PFI instrumentation showed temperatures between
+5$^\circ$C and +25$^\circ$C with a diurnal variation of only
$\sim$8$^\circ$C, thus perfectly maintaining the alignment as adjusted in the
laboratory. Several external components showed temperature variations of
typically $\pm$10$^\circ$C. They are mainly caused by the changing albedo
flux onto the instrument surfaces, resulting from changing solar elevation
and terrain underneath the gondola.

\begin{figure}[t]
 \centerline{\includegraphics[width=1.\textwidth,clip=]{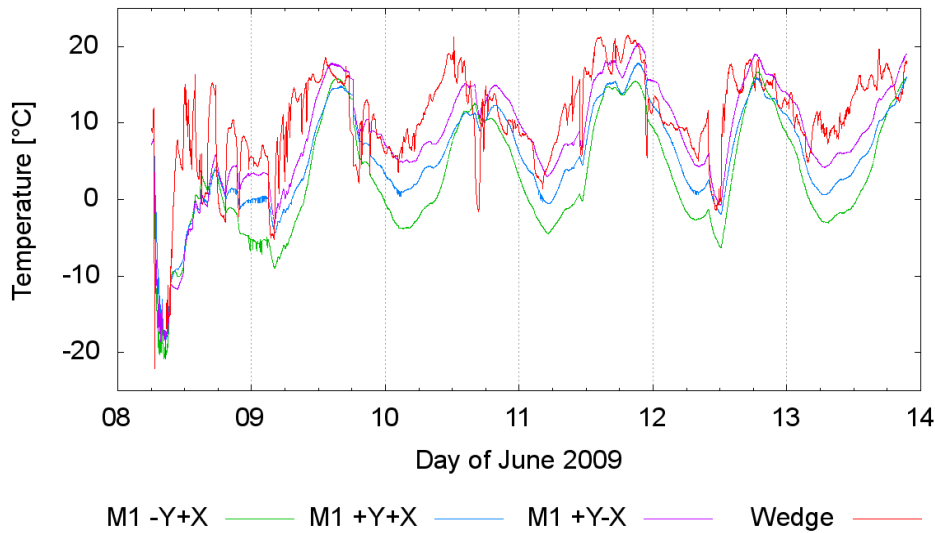}}
\caption{In-flight temperatures of 3 primary mirror sensors at several positions distributed over the aperture
and of the heat rejection wedge at the primary focus (in red)}
\label{telescope_temp}
 \end{figure}

The varying albedo flux also slightly influenced telescope component
temperatures, although the telescope remained nearly permanently oriented
towards the Sun. Figure~\ref{telescope_temp} shows a plot of three
temperature sensors mounted to the Zerodur primary mirror at various lateral
positions as well as the temperature measured at the field stop in the
primary focal plane (heat rejection wedge). The large mirror showed only
negligible temperature gradients across the aperture of less than
10$^\circ$C. More important, both mirror and heat rejection wedge had very
moderate temperatures below +22$^\circ$C throughout the flight, less than
30$^\circ$C above the free air temperature at float altitude which was
measured to be around -6$^\circ$C. This temperature difference is low enough
to exclude the risk of wavefront aberrations due to 'mirror seeing'.

{\em Wavefront Quality:} The optical performance of the instrument can be
derived from data of the CWS instrument, directly measuring the wavefront
quality in-flight, as well as from the phase diversity image analysis and
reconstruction. Both data sources prove the excellent end-to end performance
and stability of the telescope and postfocus instrumentation.

The diurnal thermo-elastic deformations and the varying orientation of the
optical system with respect to the gravity vector lead to small variations in
the relative positioning of the primary and secondary telescope mirrors.
Resulting aberrations, as defocus and coma, are measured by the wavefront
sensor CWS. They can be corrected by secondary mirror displacements. During
flight only focus was adjusted in closed-loop by continous axial
repositioning of M2 in a range of about 100~micrometer. Uncorrected, this
defocusing corresponds to a wavefront error of $\pm$0.5~$\lambda$~rms. The
CWS sensitivity and the micrometer accuracy of the M2 translation stage
helped to achieve focus accuracies below 0.01~$\lambda$~rms (Berkefeld \etal,
2010). Lateral displacements of M2 would have caused a significant image
displacement, deteriorating the co-alignment of telescope and LISS sun
sensor. The observed uncorrected coma values were less than $\lambda$/10,
small enough to be handled by the phase diversity algorithms. An additional
in-flight coma correction was therefore not performed.

Analysis of the SuFI images shows a total wavefront error in the order of
$\lambda$/5 to $\lambda$/6 rms at 300~nm, very close to the expected
performance, see Gandorfer \etal, 2010. Main contributors are astigmatism and
trefoil, most probably originating from the primary mirror, showing some
dependance on elevation angle (Hirzberger \etal, 2010a). Similar results were
obtained from phase diversity analysis of IMaX data (Martinez Pillet \etal,
2010). Differential focus between CWS and SuFI as well as CWS and IMaX was
found to be within $\lambda$/20.

{\em Pointing Performance:} The gondola pointing system worked very reliable,
keeping the instrument autonomously oriented towards the Sun. The first Sun
acquisition with aperture door opening occured 5:15~hours after launch, about
two hours later than expected. The tropopause transit had caused a temporary
freeze of the fine azimuth motor, which unfortunately wasn't switched to
rotary mode during ascent as planned for continuos heating. Once this system
was recovered, part of the commissioning time with high speed telemetry was
used to adapt gain and filter settings, optimizing the system response on the
long flight train.

The wind gusts acting on the gondola were unfortunatley strong enough to
reduce the pointing performance quite frequently. Flight data indicate that
performance was best over the free ocean between Norway and Greenland
(June~9) and less good over Greenland and the many coast lines of Northern
Canada (June~13). Higher float altitudes in general seem to be beneficial for
the pointing stability, due to the lower air density. The longest gondola
pointing within the $\pm$46 arcsec capture range of the image stabilization
system was 45 minutes. Fourtyseven time series with durations between 10 and
45 minutes were obtained with closed-loop image stabilization (see Berkefeld
\etal, 2010). In total more than 33~hours of science observation (23~\% of
the total observing time) were obtained. Details on SuFI and IMaX observing
modes, number of acquired images and and duration of time series are given in
Solanki \etal, 2010. The telescope aperture door was closed 93~times due to
pointing errors exceeding 15~arcmin. No indication of structural damage to
the telescope components in the vicinity of the primary focus was found.

Post-flight analysis of the image data gives evidence that the imaging
performance of \Sunrise is not limited by the wavefront error of the
telescope and instrument optics, but by a residual image smear of the order
of 0.03-0.04~arcsec (Berkefeld \etal, 2010), which reduces the spatial
resolution in case of SuFI to ~0.1~arcsec (Gandorfer \etal, 2010) and to
0.15-0.18~arcsec for IMaX (Martinez Pillet \etal, 2010).

Although the CWS system was able to dramatically improve the effective
pointing stability, not all components of the incoming excitation spectrum of
the balloon/gondola system could be damped out.

\begin{figure}[t]
 \centerline{\includegraphics[width=1.\textwidth,clip=]{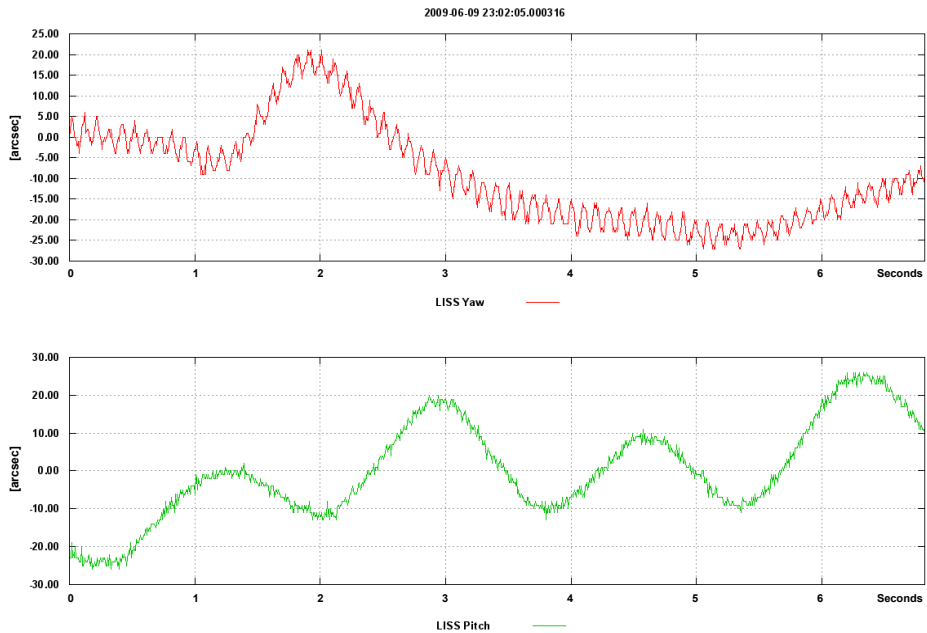}}
\caption{LISS Sun Sensor data (in arcsec) sampled with 150 Hz on June 09, 2009,
 23:02:05 UT. The length of the interval shown is 6.8 seconds, being a representative part of a $\sim$40~minutes period of closed loop pointing.
  A 10~Hz oscillation in azimuth (yaw) can clearly be seen, as well as slight pendulation in elevation (pitch)}
\label{pointing_plot}
 \end{figure}

Figure~\ref{pointing_plot} shows LISS Sun sensor data obtained on June 9,
2009 with 150~Hz sampling rate. The sample is a representative part of a
$\sim$40 minutes period, when the gondola pointing remained within the
$\pm$46~arcsec capture range of the tip-tilt mirror and both image
stabilization and telescope focus loops were closed. The data clearly show a
slight elevation pendulum motion (pitch), as well as an azimuthal (yaw)
oscillation with a frequency of ~10~Hz and several arcsec amplitude. This
oscillation most likely can be attributed to an excited resonance in the
mechanical structure of the gondola (bending mode of the solar panels),
influencing the pointing system. Amplitudes varied over the the mission and
increased by up to a factor of three towards the end.

A Fourier analysis of residual image stabilization data taken on June 12 (see
Berkefeld \etal, 2010, Figure 18) clearly shows this 10~Hz peak in the power
spectra, but also reveals additional components at frequencies above 40~Hz,
where the image stabilization is already quite limited. Both contributions
immediately died out when the azimuth motors were switched off. This can be
documented by accelerometer data taken at several positions on the gondola
structure, showing higher amplitudes at the gondola top. We therefore
currently speculate that the excitation of the 10~Hz oscillation and high
frequency jitter is connected with the azimuth drives, \eg ~by bearing
rumble. A detailed analysis of these phenomena is still ongoing, also
regarding the relative contribution to the observed image smear.

\section{Outlook}

During its 137 hours successful flight \Sunrise collected a unique data set.
The payload remained in direct sunlight the entire flight. Seeing-free
observations were possible all the time. The total observing time with image
stabilization was more than 33~hours, in which SuFI collected nearly 56.000
images in the wavelength range between 210~nm to 400~nm, being above most of
the ozone layer in the Earth's atmosphere. In the blue channels at 214~nm,
300~nm, and 312~nm, the solar surface could be observed for the first time at
high angular resolution and with unprecedented intensity constrasts (Solanki
\etal, 2010; Riethm\"{u}ller \etal, 2010; Hirzberger \etal, 2010b), thanks to
the high optical quality of the instrumentation. IMaX collected high
resolution Doppler- and magnetograms with an unprecedented combination of
angular resolution and magnetic sensitivity (Mart\'inez Pillet \etal, 2010).

The Sun was extremely quiet during the 2009 science flight. Although the
observation of quiet solar granulation is in itself an important scientific
aspect of \Sunrise, a more complete sampling of the different aspects of
solar surface magnetism is needed to finally help us understanding the
underlying physical roots of solar activity. Since the \Sunrise
instrumentation showed an excellent performance during flight and survived
the landing and recovery with only minor damage, a reflight of \Sunrise in
phases of higher solar activity has the potential of significantly advancing
our knowledge of the Sun, overcoming the current performance limitations with
relatively low additional effort.

%%%%%%%%%%%%%%%%%%%%%%%%%%%%%%%%%%%%%%%%%%%%%%%%%%%%%%%%%%%%%%%%%%%%%%%%%%%
\begin{acks}
The German contribution to \Sunrise is funded by the Bundesministerium
f\"{u}r Wirt\-schaft und Technologie through Deutsches Zentrum f\"{u}r Luft-
und Raumfahrt e.V. (DLR), Grant No. \linebreak[4] 50~OU~0401, and by the Innovationsfond of
the President of the Max Planck Society (MPG). The Spanish contribution has
been funded by the Spanish MICINN under projects ESP2006-13030-C06 and
AYA2009-14105-C06 (including European FEDER funds). The HAO contribution was
partly funded through NASA grant number NNX08AH38G. We greatly appreciate the
tremendous support provided by CSBF for this mission, especially by Danny
Ball. We would also like to thank all current and previous team members not
listed as co-authors for their valuable contribution to the project.
\end{acks}

\end{article}

\end{document}